\documentclass[useAMS,usenatbib]{mn2e}
\usepackage{graphicx}
\usepackage{amssymb}
\title[GRB X-ray flares]{Lag-luminosity relation in gamma-ray burst X-ray flares:
a direct link to the prompt emission}
\author[R. Margutti et al.]{R. Margutti$^{1,3}$\thanks{E-mail: raffaella.margutti@brera.inaf.it (RM)} ,
C. Guidorzi$^{2}$, G. Chincarini$^{1,3}$, M.G. Bernardini$^{1}$, F. Genet$^{4}$, 
\and J. Mao$^{1}$, F. Pasotti$^{1}$\\
$^{1}$INAF Osservatorio Astronomico di Brera, via Bianchi 46, Merate 23807, Italy \\
$^{2}$University of Ferrara, Physics Dept., via Saragat 1, I-44122 Ferarra, Italy\\
$^{3}$University of Milano  Bicocca, Physics Dept., P.zza della Scienza 3, Milano 20126, Italy\\
$^{4}$Racah Institute of Physics, Hebrew University of Jerusalem, Israel\\}

\begin{document}

\date{Accepted 2010 Month day. Received 2010 Month day; in original form 2010 Month day}
\pagerange{\pageref{firstpage}--\pageref{lastpage}} \pubyear{2010}
\maketitle

\label{firstpage}

\begin{abstract}
The temporal and spectral analysis of 9 bright X-ray flares out of a sample of 113 flares 
observed by \emph{Swift} reveals that the flare phenomenology is strictly
analogous to the prompt $\gamma$-ray emission: high energy flare profiles rise faster, decay faster
and peak before the low energy emission. However, flares and prompt pulses differ in one crucial aspect:
flares evolve with time. As time proceeds flares become wider, with larger peak lag, lower luminosities
and softer emission. The flare spectral peak energy $E_{\rm{p,i}}$ evolves to lower values following an exponential decay
which tracks the decay of the flare flux. The two flares with best statistics show higher than expected
isotropic energy $E_{\rm{iso}}$ and peak luminosity $L_{\rm{p,iso}}$ when compared to the 
$E_{\rm{p,i}}-E_{\rm{iso}}$ and $E_{\rm{p,i}}-L_{\rm{iso}}$ prompt correlations. $E_{\rm{p,i}}$ 
is found to correlate with $L_{\rm{iso}}$ within single flares, giving rise to a time resolved 
$E_{\rm{p,i}}(t)-L_{\rm{iso}}(t)$. Like prompt pulses, flares define a lag-luminosity 
relation: $L_{\rm{p,iso}}^{0.3-10 \,\rm{keV}}\propto t_{\rm{lag}}^{-0.95\pm0.23}$. The lag-luminosity 
is proven to be a fundamental law extending $\sim$5 
decades in time and $\sim$5 in energy. Moreover, this is direct evidence that GRB X-ray flares and prompt 
gamma-ray pulses are produced by the same mechanism. Finally we establish a flare- afterglow morphology connection: 
flares are preferentially detected superimposed to one-break or canonical X-ray afterglows.

\end{abstract}

\section{Introduction}
\label{Sec:Introduction}
The high temporal variability was one of the first properties to be attributed to the Gamma-ray burst
(GRB) prompt emission in the $\gamma$-ray energy band (\citealt{Klebesadel73}). The advent of \emph{Swift}
(\citealt{Gehrels04}) revealed that a highly variable emission characterises also
the early time X-ray afterglows in the form of erratic flares. 
This established the temporal variability as one of the key features in interpreting the GRB phenomena.   

GRB\,050502B and the X-ray flash 050406 (\citealt{Falcone06}; \citealt{Romano06b}; \citealt{Burrows05b})
provided the first examples of dramatic flaring activity superimposed to a smooth decay:
in particular, GRB\,050502B demonstrated that flares can be considerably energetic, with a 
0.3-10 keV energy release comparable to the observed prompt fluence in the 15-150 keV band. Thanks 
to the rapid re-pointing \emph{Swift} capability, it was later shown that flares are a common feature
of the early X-ray afterglows, being present in the $\sim 33\%$ 
of X-ray light-curves (\citealt{Chincarini07}, hereafter C07; \citealt{Falcone07}, hereafter F07).
On the contrary, a convincing optical flare, counterpart to a detected X-ray flare is still lacking,
suggesting that the detected optical afterglow contemporaneous to the high-energy flares is
dominated by a different emission component (see e.g. GRB\,060904B,  \citealt{Klotz08}
but see also \citealt{Greiner09} where an optical flare was probably detected but, unfortunately, 
contemporaneous X-ray coverage is lacking).

Based on the  temporal and spectral study of a statistical sample of X-ray flares within GRBs, C07 
and F07 showed that the flares share common properties and that the flare phenomenology can be described 
using averaged properties (see C07 and F07 and references therein): 
\begin{itemize}
\item The same GRB can show multiple flares (see e.g. GRB\,051117A which contains a minimum of 11 
  structures in the first 1 ks of observation);
\item The underlying continuum is consistent with having the same slope before and after the flare,
  suggesting that  flares constitute a separate component in addition to the observed continuum;
\item Each flare determines a flux enhancement evaluated at the peak time $\Delta F/ F$ between $\sim1$
  and $\sim1000$, with a fluence that competes in some cases (e.g. GRB\,050502B) with the prompt 
  $\gamma$-ray fluence. The average flare fluence is $\sim 10$\% the 15-150 keV prompt fluence;
  \item Flares are sharp structures, with $\Delta t/t \sim 0.1$, a fast rise and a slower decay; 
  \item Each flare determines a hardening during the rise time and a softening during the 
  decay time (F07), reminiscent of the prompt emission (e.g. \citealt{Ford95}): the result is a hardness ratio 
  curve that mimics the flare profile (see e.g. GRB\,051117A, \citealt{Goad07}, their figure 9). In this sense 
  flares are spectrally harder than the underlying continuum;
\item The spectrum of a consistent fraction of flares is better fitted by a Band (\citealt{Band93}) model, 
  similarly to prompt emission pulses (see e.g. \citealt{Kaneko06}). The flare spectral peak energy is likely to be
  in the soft X-ray range (a few keV). The spectrum evolves with time as testified 
  by the hardness ratio curve and by accurate spectral modelling. During the decay time a clear softening is 
  detected (e.g. \citealt{Krimm07}; \citealt{Godet07});
\item There is no correlation between the number of flares and the number of prompt emission pulses;
\item The vast majority of flares are concentrated in the first 1 ks after trigger.  However, late-time 
  flares ($t_{\rm{peak}}\sim 10^5-10^6$ s) are present as well: flares are not confined to the steep decay phase, 
  but can happen during the plateau and  the normal decay phases. Their temporal properties are consistent with 
  those of early flares (\citealt{Curran08}), even if their lower brightness prevents a detailed
  comparison with the entire set of early time flare properties (this is especially true from the spectral point of view);
\item Flares happen both in low-z and high-z environments: the record holder GRB\,090423 at z$\sim8.2$
  (\citealt{Salvaterra09}; \citealt{Tanvir09}) shows a prominent flare with standard properties when compared to the sample
  of X-ray flares of \cite{Chincarini10} (C10, hereafter);
\item Flares  have been detected both in hard and soft events such as X-Ray Flashes  (e.g. XRF\,050406);
\item Variability has also been  detected in the X-ray afterglows of \emph{short} GRBs (GRB with a 
  prompt emission duration $T_{90}<2$ s, \citealt{Kouveliotou93}). However, given the lower brightness associated
  to these events it is still unclear if  what is currently identified as a short GRB flare emission, quantitatively shares 
  the very same properties as the population of flares detected in \emph{long} GRBs. GRB\,050724  (\citealt{Barthelmy05b}) 
  constitutes a good example of short GRB with late-time variability.
\end{itemize}

From the systematic study of 113 flares in the XRT 0.3-10 keV energy band, as well as in 4 sub-energy bands, C10 demonstrated that:
\begin{itemize}
\item Flares are asymmetric with an average asymmetry parameter similar to the prompt emission value; no flare is found
  rising slower than decaying;
\item The flare width evolves linearly with time $w\propto 0.2\, t_{\rm{peak}}$. This is a key point which 
  clearly distinguishes the flares from the prompt emission, where no evolution of the pulse-width has ever been found
  (see e.g. \citealt{Ramirez00});
\item The width evolution is the result of the linear evolution of both the rise and the decay times: $t_{\rm{r}}\propto0.06\, t_{\rm{peak}}$;
	$t_{\rm{d}}\propto 0.14\,t_{\rm{peak}}$.
\item The asymmetry does not evolve with time. Instead the rise over decay time ratio is constant with time, 
  implying that both time scales are stretched of the same factor. Furthermore $t_{\rm{d}}\sim2\,t_{\rm{r}}$. Flares are \emph{self-similar}
  in time.
\item At high energy the flares are sharper with shorter duration:  $w\propto E^{-0.5}$. Prompt pulses share the
  same property, with a similar dependence on the energy band (\citealt{Fenimore95}; \citealt{Norris96});
\item The flare peak luminosity decreases with time. Accounting for the sample variance
 the best fit relation reads: $L_{\rm{peak}}\propto t_{\rm{peak}} ^{-2.7\pm0.5}$. The average flare luminosity 
  declines as a power-law in time $<L>\propto t^{-1.5}$ (\citealt{Lazzati08});
\item The isotropic 0.3-10 keV flare energy distribution is a log normal peaked at $\sim 10^{51}$ erg (this can be 
  viewed as the typical flare isotropic energy). The distribution further shows a hint of bimodality. 
\item In multiple-flare GRBs, the flares follow a softening trend which causes later time flares to be 
  softer and softer: the emitting mechanism keeps track of the previous episodes of emission. 
\end{itemize}

Starting from these pieces of evidence, C07, F07 and C10 concluded that 
in spite of the softness and width evolution, the X-ray flares and the prompt pulses are likely 
to share a common origin. However  important questions are still to be addressed: do flares follow
the entire set of temporal and spectral relations found from the analysis of prompt emission pulses?
In particular: is it possible to define a flare peak-lag? Do flares follow a lag-luminosity relation?
Is the lag-time linked to other temporal properties (e.g. the flare duration, asymmetry, etc.)? What can be said about the pulse start 
conjecture for flares (\citealt{Hakkila09})? Is it possible to quantify the evolution of the flare temporal
properties as a function of the energy band? Do the rise and decay times evolve differently with energy 
band? Is it possible to track and quantify the evolution of the spectral peak energy $E_{\rm p}$ during 
the flare emission? Is there any connection between the temporal and spectral properties of the flares?
What is the position and the track of the flares in the 
$E_{\rm{p}}-L_{\rm{iso}}$ (spectral peak energy-isotropic luminosity) plane? 
Is there any link between the flares and the underlying X-ray afterglow morphology?

This set of still open questions constitutes the major motivation for undertaking the present 
investigation. The primary goal of this paper is to model the X-ray flare profiles and 
constrain their evolution with energy to obtain parameters
that uniquely qualify the shape and spectrum of the flares and compare those
values to the well known signatures of the prompt emission pulses.
Since this is often difficult because of low statistics, overlap or non trivial estimate of the underlying
continuum, the present work concentrates on  very bright and isolated flares for which
the underlying continuum does not play a major role neither from the 
temporal point of view nor from the spectral point of view. This work is observationally driven: a critical review of 
theoretical models in the light of the present results is in preparation.

This work is organised as follows: the sample selection and data reduction are described in 
Sect. \ref{Sec:datared}. In Sect. \ref{Sec:tempan} we perform the flare temporal analysis, 
while to the spectral properties of the sample is dedicated Sect. \ref{Sec:spec}. The 
discussion follows (Sect. \ref{Sec:disc}). Conclusions are drawn in Sect. \ref{Sec:conc}.

The phenomenology of the  different GRBs is presented in the observer frame unless 
otherwise stated.  The 0.3-10 keV energy band is adopted unless specified.
The zero time is assumed to be the trigger time.  
The convention $F(\nu,t)\propto \nu^{-\beta}t^{-\alpha}$
is followed, where $\beta$ is the spectral index, related to the photon
index $\Gamma$ by $\Gamma=\beta+1$.
All the quoted uncertainties are given at 68\% confidence
level (c.l.): a warning is added if it is not the case.
Standard cosmological quantities have been adopted: 
$H_{0}=70\,\rm{Km\,s^{-1}\,Mpc^{-1}}$, $\Omega_{\Lambda }=0.7$, 
$\Omega_{\rm{M}}=0.3$.\\
\section{Sample selection and data reduction}
\label{Sec:datared}
We select the brightest, isolated flares detected by the X-Ray Telescope (XRT, \citealt{Burrows05})
on board \emph{Swift} (\citealt{Gehrels04}) in the time period April 2005 - January 2010. The sample
comprises 9 flares detected in 8 different GRBs\footnote{GRB\,050502B is not included since the 
bright flare detected by the XRT is possibly due to the superposition of more than one structure 
(\citealt{Burrows05b}; C10)}: GRB\,050822, GRB\,060418, GRB\,060526, 
GRB\,060904B, GRB\,060929, GRB\,070520B, GRB\,070704 and GRB\,090621A\footnote{Note that
for GRB\,090621A, the Burst Alert Telescope (BAT, \citealt{Barthelmy05}) triggered on the precursor 
(\citealt{Curran09}). } (see Table \ref{Tab:bestfit}).  The required isolation 
allows us to constrain the flares temporal and spectral properties with high accuracy; their
brightness assures the possibility to analyse in detail the evolution of their temporal properties in different
energy bands inside the 0.3-10 keV of the XRT; at the same time, this requirement guarantees
that the partial ignorance of the underlying continuum properties has a negligible impact on our
conclusions, both from the spectral and from the temporal point of view.

XRT data have been processed with the \textsc{heasoft} package v. 6.6.1 and corresponding 
calibration files: standard filtering and screening criteria have been applied.
\emph{Swift}-XRT  is designed to acquire data using different observing modes
to minimise the presence of pile-up. When the source is 
brighter than a few $\rm{count\,s^{-1}}$, the data are acquired in Windowed Timing (WT) mode.
In this case we applied the  standard pile-up corrections following the prescriptions of  
\cite{Romano06} when necessary. For lower count-rates the spacecraft 
automatically switches to the Photon Counting (PC) mode to follow the fading of the source. 
The events are extracted from circular regions centred at the afterglow position with 
progressively smaller radii to assure the best signal to noise ratio (SN). 
When the PC data suffered from pile-up, we extracted the source events in an annulus
whose inner radius  is derived comparing the observed to the nominal point spread function 
(PSF, \citealt{Moretti05}; \citealt{Vaughan06}). The background is estimated from a source-free 
portion of the sky and then subtracted. The background subtracted, PSF and vignetting 
corrected light-curves were then re-binned 
so as to assure a minimum signal-to-noise (SN) equals to 4 for WT and PC mode data. Data coming from different orbits 
of observation were merged to build a unique data point, when necessary. This procedure was
applied to extract GRB light-curves in the nominal XRT energy range (0.3-10 keV) as well as
count-rate light-curves of the same event in 4 different sub-energy bands: 0.3-1 keV; 
1-2 keV; 2-3 keV and 3-10 keV \footnote{Note that, differently from C10
the described data reduction assures a minimum SN equal to 4 for \emph{each} data point of the 
GRB light-curves in the different energy bands. This allows a higher sensitivity which translates
into a more accurate evaluation of the best fit parameters.}.
\section{Flare temporal analysis}
\label{Sec:tempan}

\begin{figure}
\vskip -0.0 true cm
\centering
    \includegraphics[scale=0.43]{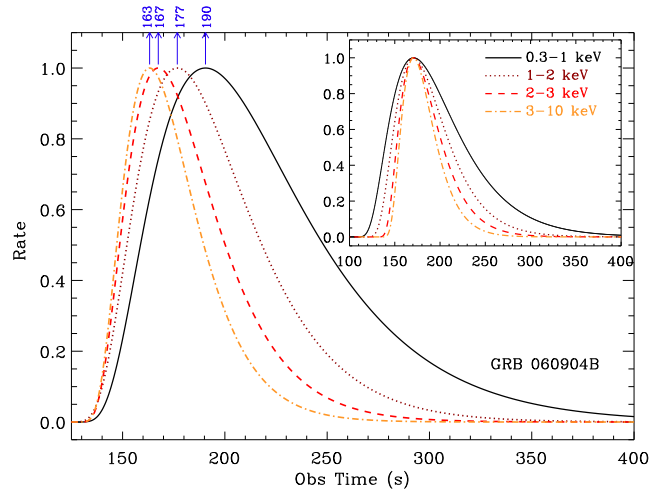}
      \caption{Best fit profile of the flare detected in GRB\,060904B in different XRT energy bands.
      The thick line, dotted line, dashed line and dot-dashed line refer to the flare profile 
      observed in the 0.3-1 keV, 1-2 keV, 2-3 keV and 3-10 keV energy band, respectively. The 
      profiles have been re-normalised for the sake of clarity. The blue arrows point to the 
      flare peak times in the different energy bands. Softer energy bands profiles peak later.
      \emph{Inset:} The re-normalised best fit
      profiles have been aligned at $t_{\rm{peak}}=170.71$ s, peak time of the flare in the 0.3-10 keV energy
      band as derived from the best fit parameters of Table \ref{Tab:bestfit}. Softest profiles are wider.  }
\label{Fig:060904Bprofile}
\end{figure}
\begin{figure*}
\vskip -0.0 true cm
\centering
    \includegraphics[scale=0.85]{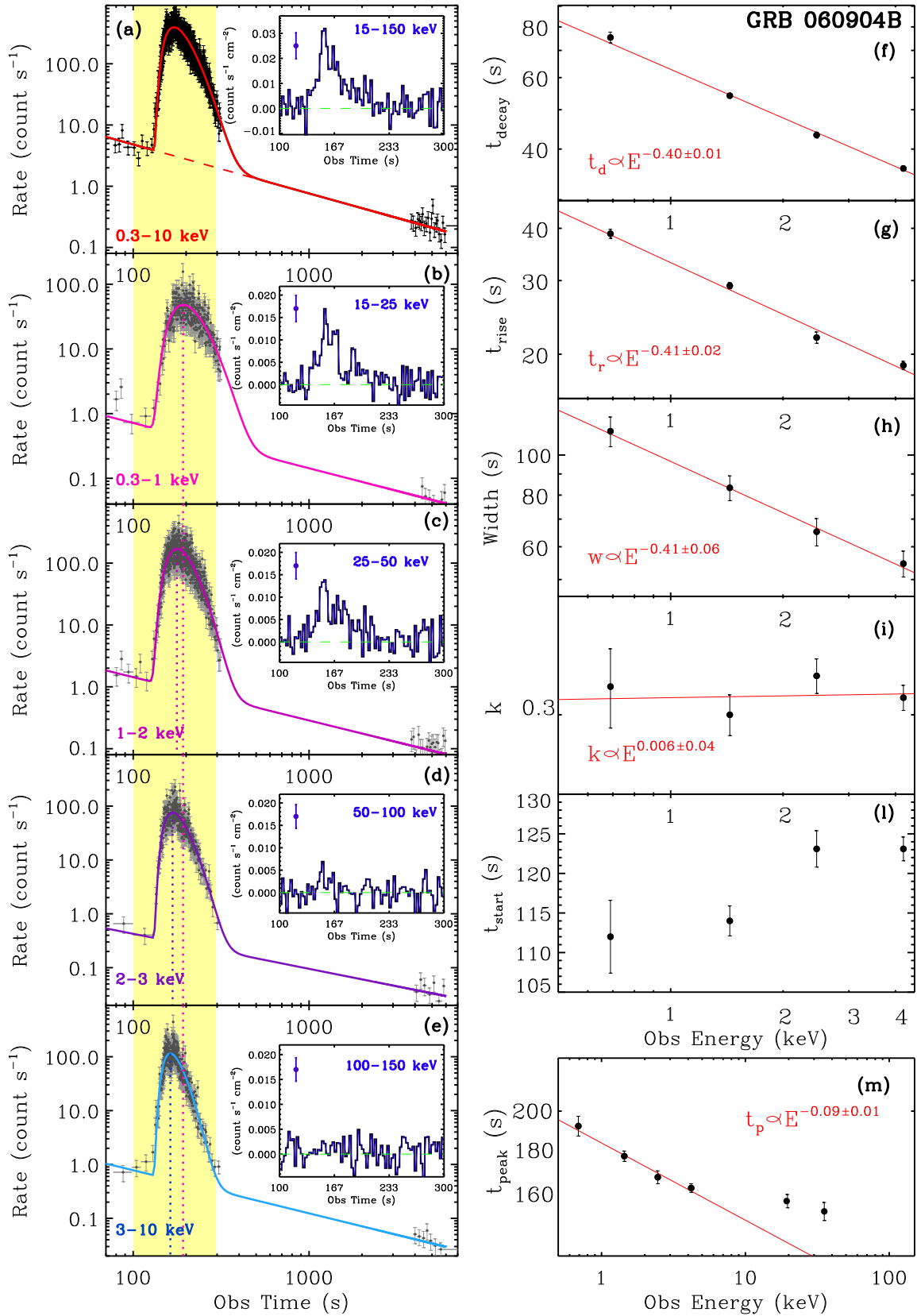}
      \caption{Panels \emph{(a)} through \emph{(e)}: GRB\,060904B flare best fit profile in the total 0.3-10 keV
        XRT energy band and in the 4 channels. \emph{Insets}: BAT signal contemporaneous to the flare emission detected in the 15-150 keV, 15-25 keV,
      25-50 keV, 50-100 keV and 100-150 keV energy bands. The typical uncertainty affecting the BAT data is also
      shown. Panels \emph{(f)} through \emph{(m)}: observed evolution of the flare decay time \emph{(f)}, rise time \emph{(g)},
      width \emph{(h)}, asymmetry \emph{(i)}, start-time \emph{(l)} and peak time \emph{(m)} with observed energy band.
      The best fit power-law relation describing the evolution of each parameter is drawn with a red solid line and explicitly
      written in each panel. The fit reported in panel \emph{(m)} concerns the XRT data only.}
\label{Fig:060904B}
\end{figure*}

Each flare profile in each energy band is modelled using the \cite{Norris05} (hereafter N05) 
profile. This choice allows a direct comparison between the properties of the X-ray flares and 
of the wide, long-lag prompt pulses analysed by N05 (see Sec. \ref{Sec:disc}). The N05 profile reads:
\begin{equation}
	\label{Eq:norris05}
	I(t)=A\lambda \exp[-\frac{\tau_1}{(t-t_{\rm{s}})}-\frac{(t-t_{\rm{s}})}{\tau_2}]\,\,\,\rm{for}\,\,\,t>t_{\rm{s}}
\end{equation}
where $\lambda=\exp(2\mu)$ and $\mu=(\tau_1/\tau_2)^{1/2}$. $t_s$ is the pulse onset time.
The parameters are defined following N05. In particular,  the intensity peaks at:
\begin{equation}
	\label{Eq:tpeak}
	t_{\rm{peak}}=t_s+(\tau_1 \tau_2)^{1/2}
\end{equation}
The pulse width is measured between the two $1/e$ intensity points and consequently 
reads:
\begin{equation}
	\label{Eq:width}
	w=\tau_2(1+4\mu)^{1/2}=t_{\rm{r}}+t_{\rm{d}}
\end{equation}
The asymmetry is defined as:
\begin{equation}
	\label{Eq:asymmetry}
	k=(1+4\mu)^{-1/2}=\frac{t_{\rm{d}}-t_{\rm{r}}}{t_{\rm{d}}+t_{\rm{r}}}
\end{equation}
where $t_d$ and $t_r$ are the  $1/e$ decaying and rising
times, respectively:
\begin{equation}
	\label{Eq:trisetdecay}
	t_{\rm{d,r}}=\frac{1}{2}\tau_2[(1+4\mu)^{1/2}\pm 1]
\end{equation}
We account for the entire covariance matrix during the error propagation procedure.
The continuum underlying the flare emission is estimated from data acquired before and after the flare and is
modelled using a power-law or a broken power-law: the parameters related to the continuum 
have been left free to vary in a first fit and then frozen to their best fit values in a second fit.
The best fit parameters and related quantities are reported in Table \ref{Tab:bestfit}.
Figure \ref{Fig:060904Bprofile} shows the best fit profiles in the different XRT band-passes
obtained for the flare detected in GRB\,060904B, taken as an example. A general trend can be seen 
for the flare to be wider and peak later at lower energies. A complete summary of the results 
obtained from the fitting procedure in different energy bands is portrayed in Fig. 
\ref{Fig:060904B}: the detected evolution of the different parameters with energy 
(panels \emph{f} to \emph{m}) will be extensively treated in SubSec. \ref{SubSec:parevolfreq} and
\ref{SubSec:lag}. The flare detected in GRB\,060904B is here shown as an example.

\subsection{Best fit parameters correlations}
\begin{figure*}
\vskip -0.0 true cm
\centering
    \includegraphics[scale=0.8]{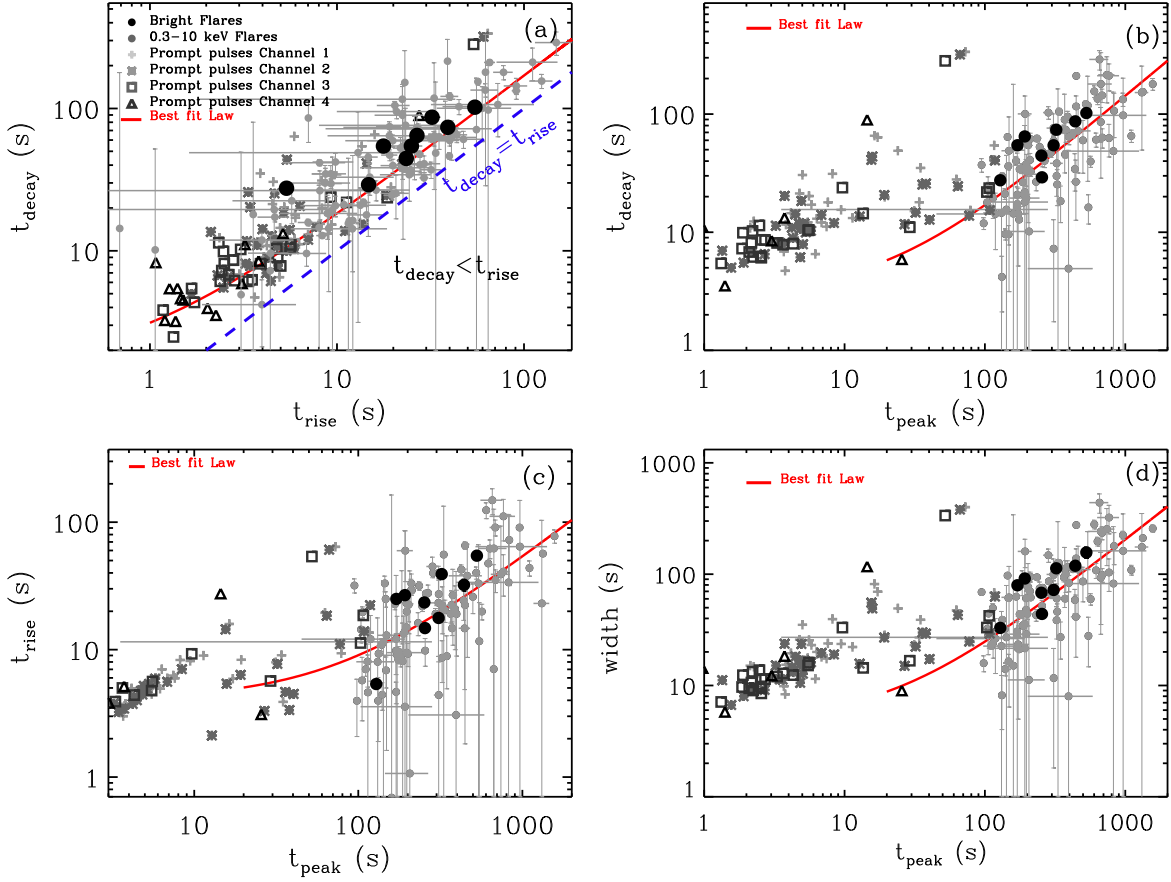}
      \caption{Scatter plot of  the temporal parameters describing the flare profile as derived from the
    best fit values reported in Table \ref{Tab:bestfit}.  \emph{(a)}: decay vs. rise time  for our sample
    of bright flares (black dots), for the sample of flares analysed by C10 (grey dots) compared
    to the values found by N05 for a sample of long-lag, wide prompt pulses detected by BATSE. Blue
    dashed line: locus of the points for which $t_{\rm{r}}=t_{\rm{d}}$. Red solid line: best fit law 
    for the three samples: $t_{\rm{d}}=(1.1 \pm 0.1)+(1.9\pm 0.1)t_{\rm{r}}$. \emph{(b)}: decay 
    vs. peak time for the three samples described for panel \emph{(a)}. Red solid line: linear best fit relation found
    for the flare samples: $t_{\rm{d}}=(3.0\pm0.4)+(0.14\pm0.02)t_{\rm{peak}}$. \emph{(c)}: rise  vs. peak time. 
    The best fit relation found for flares is  marked with a red solid line: 
    $t_{\rm{r}}=(4.0\pm0.3)+(0.06\pm0.01)t_{\rm{peak}}$. \emph{(d)}: width vs. peak time. Red solid line:
    flare best fit relation: $w=(4.8\pm0.6)+(0.20\pm0.02)t_{\rm{peak}}$.}
\label{Fig:Obsbestfitpar1}
\end{figure*}

\begin{figure*}
\vskip -0.0 true cm
\centering
    \includegraphics[scale=0.8]{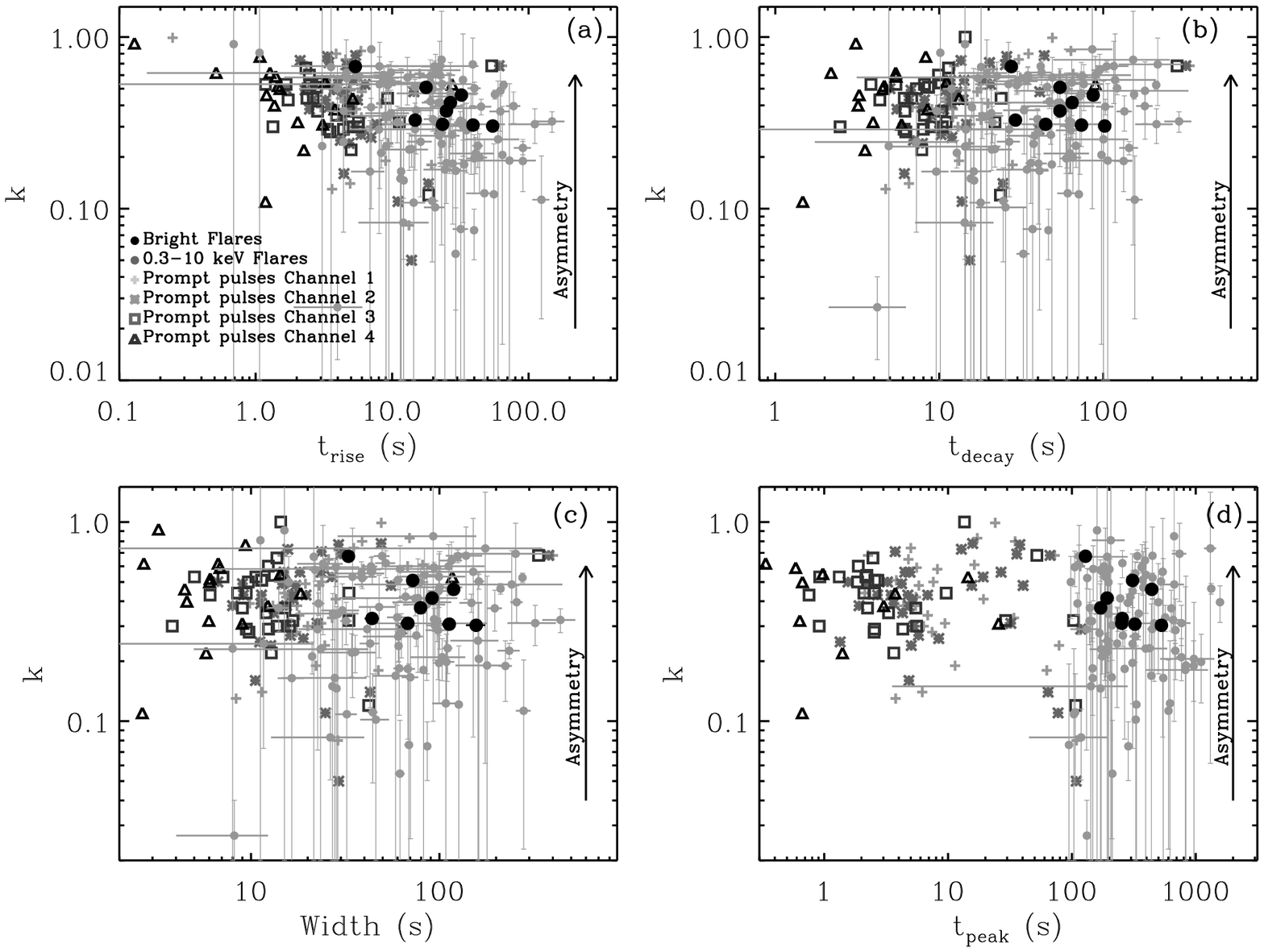}
      \caption{Asymmetry vs.  rise time (panel \emph{(a)}), decay time  (panel \emph{(b)}), width  (panel \emph{(c)}) and
      peak time (panel \emph{(d)}) for our sample of bright X-ray flares (black dots), the sample of flares from 
      C10 (grey dots) and for a sample of long-lag, wide prompt pulses from N05. In each plot, the arrow
      points to the direction of growing asymmetry.}
\label{Fig:Obsbestfitpar2}
\end{figure*}

Figures \ref{Fig:Obsbestfitpar1} and \ref{Fig:Obsbestfitpar2} show the correlations or the 
lack thereof, for the quantities derived from the best fit parameters of Table \ref{Tab:bestfit}
superimposed to the results obtained by N05 for a sample of long-lag wide prompt GRB pulses
and by C10 for a sample of 113 early X-ray flares. 
In these plots we see that bright flares obey the set of correlations found for the
complete X-ray flare catalogue: the rise time is linearly correlated with the decay time, 
and both linearly evolve with time. This gives rise to a flare width that linearly 
grows with time (see Fig. \ref{Fig:Obsbestfitpar1}). 

Figure \ref{Fig:Obsbestfitpar2} depicts instead the relation between the asymmetry $k$ and
the temporal parameters describing the flare profiles and demonstrates that it is not possible
to distinguish the bright from the main sample on the basis of their asymmetry.
Furthermore, the asymmetry seems to be independent of the other parameters 
(Fig. \ref{Fig:Obsbestfitpar2}).

\subsection{Best fit parameters evolution with energy}
\label{SubSec:parevolfreq}
\begin{figure}
\vskip -0.0 true cm
\centering
    \includegraphics[scale=0.83]{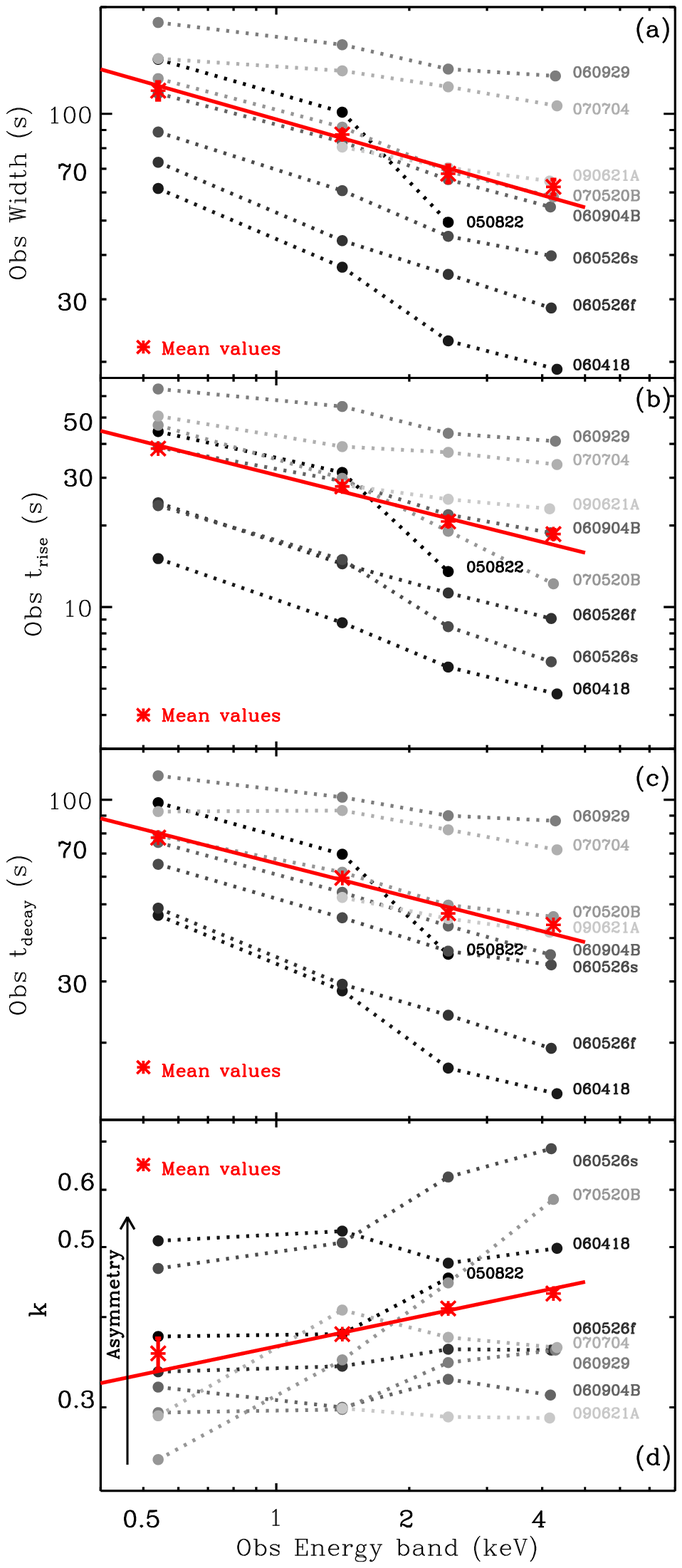}
      \caption{Evolution of the width \emph{(a)}, rise time \emph{(b)},  decay time \emph{(c)}
      and asymmetry \emph{(d)} with observed energy band obtained for our sample of bright X-ray flares. 
      Results coming from the fit of a particular flare in different energy bands are indicated by 
      dots and linked by a dotted line. Red stars:  mean values derived from the entire sample. Red
      solid line: best fit power-law model:
       \emph{(a)} $w=10^{1.98\pm0.01}E^{-0.32\pm0.01}$;
      \emph{(b)} $t_r=10^{1.48\pm0.01}E^{-0.37\pm0.01}$;
      \emph{(c)} $t_d=10^{1.81\pm0.01}E^{-0.29\pm0.01}$;
      \emph{(d)} $k=10^{-0.44\pm0.01}E^{0.11\pm0.01}$.}
\label{Fig:par_energy}
\end{figure}

The well known trend of narrower pulses at higher energy is apparent in Fig. 
\ref{Fig:060904Bprofile} for the flare detected in GRB\,060904B. This trend is quantified in Fig.
\ref{Fig:par_energy} for the entire sample: the narrowing with energy follows a power-law behaviour\footnote{
For the 0.3-1 keV, 1-2 keV, 2-3 keV energy bands, we define an effective energy 
$E_{\rm{eff}}=\frac{\int^{E_1}_{E_2} E f(E)RM(E)d(E)}{\int^{E_1}_{E_2} f(E)RM(E)d(E)}$
where $RM$ stands for the XRT response matrix, $f(E)\sim E^{-1}$, as found by C10
for the average flare spectral energy distribution. For the hardest band
the best fit spectrum of each flare is used.}:
$w\propto E^{-0.32\pm0.01}$.
The same is true for the rise and decay times: $t_{\rm{r}}\propto E^{-0.37\pm0.01}$;
$t_{\rm{d}}\propto E^{-0.29\pm0.01}$. The width evolution is the combined result of 
the evolution of these two time scales and its evolution is consequently
associated to a power-law index which is between 0.29 and 0.37.
Notably the rise time shows a steeper dependence on the energy band than the 
decay time. As a result, the asymmetry is slightly dependent on energy:
flares are on average more asymmetric at higher energies.  
Flare profiles peak later at lower energies (Fig. \ref{Fig:060904B}, panel \emph{m}); on the contrary, 
it seems that high energy profiles start their rise later (Fig. \ref{Fig:060904B}, panel \emph{l}):
while the flare in GRB\,060904B is here portrayed as an example, the same is true for the entire sample
as testified by the results of Table \ref{Tab:energyevol}. 

Figure \ref{Fig:par_energy} also demonstrates that the energy dependence of the
various parameters varies from one flare to another: the average relations describe
the general behaviour of the parameters with energy band, but the evolution
in a particular flare can be markedly different. This statement 
is quantified in Table \ref{Tab:energyevol}, where the best fit power-law index 
describing the behaviour of the temporal parameters with energy band is listed.
The best fit relations are portrayed in Fig. \ref{Fig:060904B}, panels \emph{(f)}
through \emph{(m)} for GRB\,060904B. 
How this different evolution rate relates to the spectral properties of the
flares is investigated in Sec. \ref{Sec:spec}.

\begin{figure}
\vskip -0.0 true cm
\centering
    \includegraphics[scale=0.5]{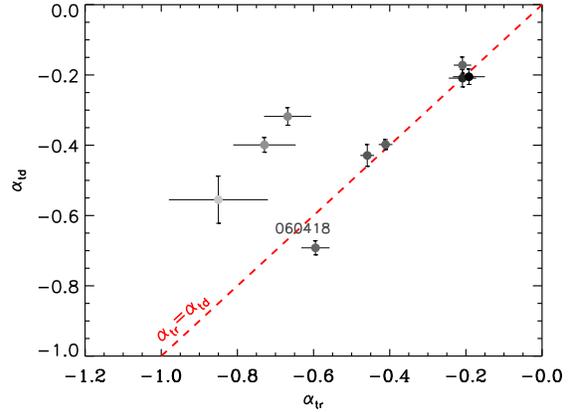}
      \caption{Best fit power-law index describing the evolution of the decay time with observed energy 
      as a function of the best fit power-law index found for the evolution of the rise time with energy. 
      Red dashed line: locus of the points for which $\alpha_{td}=\alpha_{tr}$.  Darker points are 
      associated to spectrally harder flares.}
\label{Fig:trisetdecayslope}
\end{figure}

\begin{table*}
 \centering
  \begin{minipage}{180mm}
  \caption{Evolution of the flare profile with energy band. The rise time, decay time, peak time, width and
  asymmetry have been fitted using a power-law model in energy: $y\propto E^{\alpha}$. From left to right:
  power-law index associated to the evolution of the following parameters: rise time $\alpha_{tr}$, decay time  $\alpha_{td}$, 
  peak time  $\alpha_{tp}$, width $\alpha_{w}$ and asymmetry  $\alpha_{k}$.}
\label{Tab:energyevol}
  \begin{tabular}{llllll}
  \hline
   GRB & $\alpha_{tr}$& $\alpha_{td}$& $\alpha_{tp}$&$\alpha_{w}$&$\alpha_{k}$ \\
   \hline
 050822    &$  -0.85   \pm  0.13     $&$   -0.56   \pm      0.07     $&$          -0.11    \pm       0.03    $&$       -0.83      \pm    0.10   $&$            0.12   \pm        0.04     $\\
 060418    &$  -0.60   \pm  0.04     $&$   -0.69  \pm       0.02    $&$          -0.021 \pm        0.005    $&$       -0.68      \pm    0.02   $&$           -0.034 \pm        0.009   $\\
 060526f\footnote{First flare detected by the XRT;}   &$  -0.46    \pm 0.02     $&$   -0.43  \pm       0.03     $&$         -0.029  \pm       0.001    $&$       -0.44       \pm   0.02   $&$            0.040  \pm        0.005   $\\       
 060526s\footnote{Second flare detected by the XRT.}  & $ -0.73   \pm  0.08     $&$    -0.40  \pm       0.02     $&$         -0.042 \pm        0.006    $&$       -0.43      \pm    0.01   $&$           0.24    \pm        0.01   $\\    
 060904B  &$  -0.41   \pm  0.02    $&$     -0.40  \pm       0.02    $&$          -0.090    \pm       0.014  $&$       -0.41     \pm     0.06  $&$            0.0060 \pm      0.039      $\\
 060929    &$  -0.21   \pm  0.02    $&$      -0.17  \pm       0.02      $&$        -0.050  \pm       0.006   $&$        -0.20     \pm     0.04  $&$           0.090   \pm      0.018      $\\         
 070520B  & $ -0.67   \pm  0.06    $&$      -0.32  \pm       0.03    $&$          -0.063    \pm      0.025   $&$       -0.41     \pm     0.07  $&$           0.48    \pm      0.05     $\\      
 070704    &$  -0.21    \pm 0.04     $&$     -0.21   \pm      0.03      $&$        -0.033    \pm      0.011   $&$       -0.18      \pm     0.04  $&$          0.039   \pm     0.021      $\\
 090621A  & $ -0.19    \pm 0.04    $&$      -0.21  \pm       0.02     $&$         -0.029   \pm       0.021   $&$       -0.21      \pm     0.18  $&$         -0.028   \pm      0.057      $\\   
 \hline
 \end{tabular}
  \end{minipage}
\end{table*}
\subsection{Pulse peak lag}
\label{SubSec:lag}

\begin{figure*}
\vskip -0.0 true cm
\centering
    \includegraphics[scale=0.75]{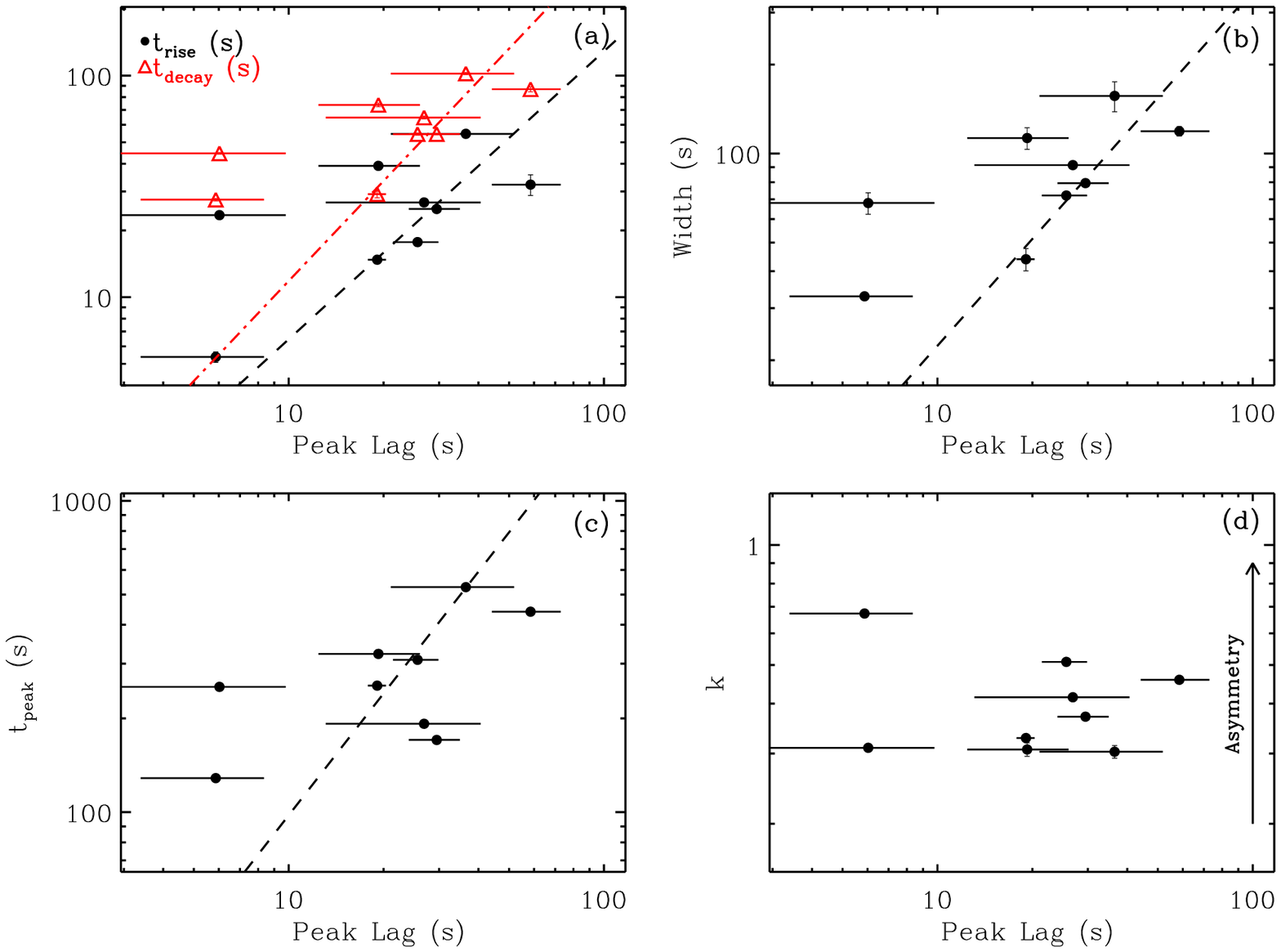}
      \caption{\emph{(a)}: Bright flares rise (red triangles) and decay times (black dots) vs. peak lag.
      The red dot-dashed line and black dashed line indicate the best fit relations: 
      	$t_{\rm{r}}=10^{-0.49\pm0.28} (t_{\rm{lag}})^{1.3\pm0.21}$;
             $t_{\rm{d}}=10^{-0.31\pm0.32} (t_{\rm{lag}})^{1.5\pm0.24}$.
       \emph{(b)}: Width vs. peak lag. The dashed line indicates the best fit relation: 
       $w=10^{0.15\pm0.27} (t_{\rm{lag}})^{1.2\pm0.20}$.
      \emph{(c)}: peak time vs. peak lag. The dashed line indicates the best fit relation: 
      $t_{\rm{peak}}=10^{0.69\pm0.28} (t_{\rm{lag}})^{1.2\pm0.21}$.  \emph{(d)}: scatter plot of
      the asymmetry vs. peak lag. No relation is apparent. The arrow points to the direction of growing
      asymmetry.}
\label{Fig:Obslag_corr}
\end{figure*}

A commonly used parameter related to the spectral evolution of the pulses is the time-lag
between two energy channels. In this work we always refer to the flare \emph{peak lag} which is defined as
the difference between the peak of  the lowest and highest energy profiles measured by the XRT
as derived from the best fit parameters of Table \ref{Tab:bestfit}. 
Taking the flare detected in GRB\,060904B as an example, from Fig. \ref{Fig:060904Bprofile}  a clear 
tendency is apparent for the peak in the harder channel
to lead that in the softer channel: this translates into a measurable positive pulse peak lag.  The same is true
for the entire sample of flares. The evolution of the flare peak times as a function of energy  in the XRT bandpass
is reasonably well represented by a power-law, as testified by Table \ref{Tab:energyevol}  and by Fig. 
\ref{Fig:060904B}. Again, different flares have different
rates of evolution of their peak times in fixed energy band-passes.  

We scanned the BAT data looking for detections of the bright  flares  in the 15-150 keV  energy range: a
peak finding algorithm is run. The software is written to automatically scan the observations using all
the possible re-binning time scales: the central time of the bin which maximises the detection is used 
as flare peak time.  The flare peak times in the $\gamma$-ray band have been added to Fig. \ref{Fig:060904B},
panel \emph{(m)} when a $5\sigma$ detection was found. As appears from this figure,
the flare peaks at high energy are not consistent with the extrapolation of the law deduced from X-ray 
data alone. The same is true for the flares in GRB\,060418, GRB\,060526 and GRB\,070704 which have  a BAT
detection contemporaneous to the flare emission. With reference to GRB\,060904B,  it is notable that the highest rate of evolution of $t_{\rm{peak}}$ 
is found for band-passes which are crossed by the spectral peak energy $E_{\rm{p}}$ during the flare emission
(see SubSec. \ref{SubSec:specmodeling}).

The presence of correlations between the peak-lag and the flare shape parameters is investigated in Fig.
\ref{Fig:Obslag_corr}. The lag is found to be positively correlated with both
the rise and the decay times (panel \emph{a}), giving rise to the  width-lag correlation portrayed in panel \emph{b}:
the wider the flare, the higher the lag value.
The same is true for the flare peak time: later flares are associated to larger time lags. 
\section{Flare spectral analysis}
\label{Sec:spec}

\begin{figure}
\vskip -0.0 true cm
\centering
    \includegraphics[scale=0.55]{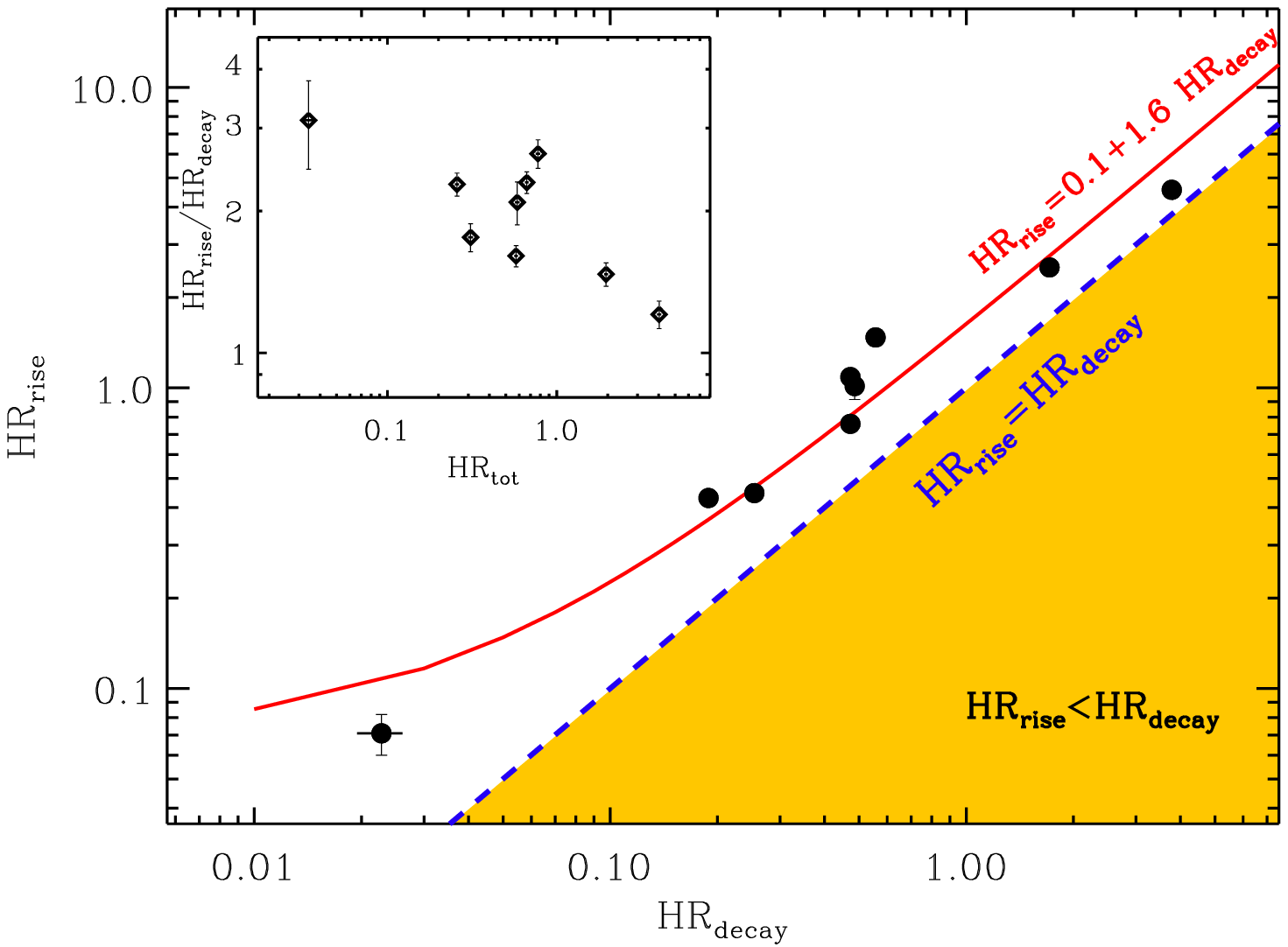}
      \caption{Hardness ratio of each flare during the rise time vs. hardness ratio during the decay time:
       the flare spectrum is harder during the rise time. Red solid line: best fit linear relation:
      $\rm{HR_{rise}}=(0.07\pm0.01)+(1.6\pm0.1) \rm{HR_{decay}}$. \emph{Inset:} rise time over decay time
      hardness ratio vs. overall spectral hardness of the flares: a trend is apparent for harder flares to show 
      more similar hardness ratios during the rise and decay portions of X-ray the light-curves. }
\label{Fig:hr_rise_decay}
\end{figure}

The flare emission has been proven to undergo a strong spectral evolution with the hardness ratio
tracking the flare profile (see e.g. \citealt{Goad07}). This results in flares profiles on average harder
than the underlying continuum.
In this section we quantify the spectral 
properties of each flare of our sample using the hardness ratio analysis (Subsec. \ref{SubSec:HRan}).  
A link between the  temporal and spectral properties is established in  Subsec. 
\ref{SubSec:evolHR}, while a proper spectral modelling is performed for the two flares with the best statistics, measured 
redshift and BAT detection in Subsec. \ref{SubSec:specmodeling}.  
\subsection{Hardness ratio analysis}
\label{SubSec:HRan}
We define the total hardness ratio as the ratio between the re-constructed counts in the 0.3-2 keV and
2-10 keV energy bands, calculated in the time interval $t_{\rm{i}}$ through $t_{\rm{f}}$, where 
$t_{\rm{i}}=t_{\rm{peak}}-t_{\rm{r}}$ and $t_{\rm{f}}=t_{\rm{peak}}+t_{\rm{d}}$ . $t_{\rm{peak}}$, $t_{\rm{r}}$ and $t_{\rm{d}}$ are derived from the 
best fit parameters of the 0.3-10 keV profile:
\begin{equation}
\rm{HR_{tot}}=\frac{Counts(2-10 \,keV)}{Counts(0.3-2\,keV)}\Big|_{t_i}^{t_f}
\end{equation}
During the rise and the decay of each flare, the hardness ratio is defined as follows: 
\begin{equation}
\rm{HR_{rise}}=\frac{Counts(2-10\,keV)}{Counts(0.3-2\,keV)}\Big|_{t_i}^{t_{peak}}
\end{equation}
\begin{equation}
\rm{HR_{decay}}=\frac{Counts(2-10\,keV)}{Counts(0.3-2\,keV)}\Big|_{t_{peak}}^{t_f}
\end{equation}

Figure \ref{Fig:hr_rise_decay} clearly shows that flares are harder during the rise time
and softer during the decay time.  The inset of this figure
illustrates the presence of a trend: the harder the overall profile, the lower is the spectral ratio
between the rise and the decay portions of a flare.  A lower degree of spectral evolution is 
detected for spectrally harder flares, in agreement with the findings of Subsect. \ref{SubSec:evolHR}.
Note however that the \emph{difference} between the $\rm{HR_{rise}}$ and $\rm{HR_{decay}}$ is higher for 
harder $\rm{HR_{tot}}$.

\begin{figure*}
\vskip -0.0 true cm
\centering
    \includegraphics[scale=0.8]{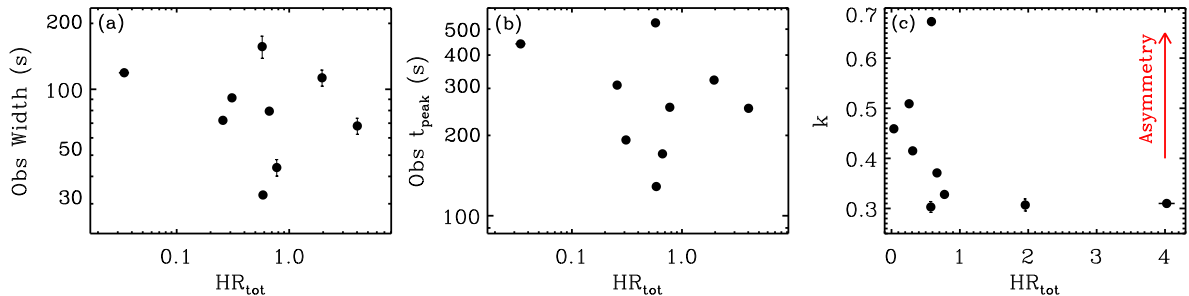}
      \caption{Best fit observed width \emph{(a)}, peak time \emph{(b)} and asymmetry \emph{(c)} as a
       function of the total hardness ratio. The red arrow in panel \emph{(c)} points to the direction of
       higher asymmetry.}
\label{Fig:par_hr}
\end{figure*}

The connection between the temporal and spectral properties is made in Fig. \ref{Fig:par_hr}. 
Panel \emph{(c)} clearly shows that hard flares are less asymmetric.

\subsection{Rate of evolution of a flare profile with energy vs. HR}
\label{SubSec:evolHR}

\begin{figure}
\vskip -0.0 true cm
\centering
    \includegraphics[scale=0.8]{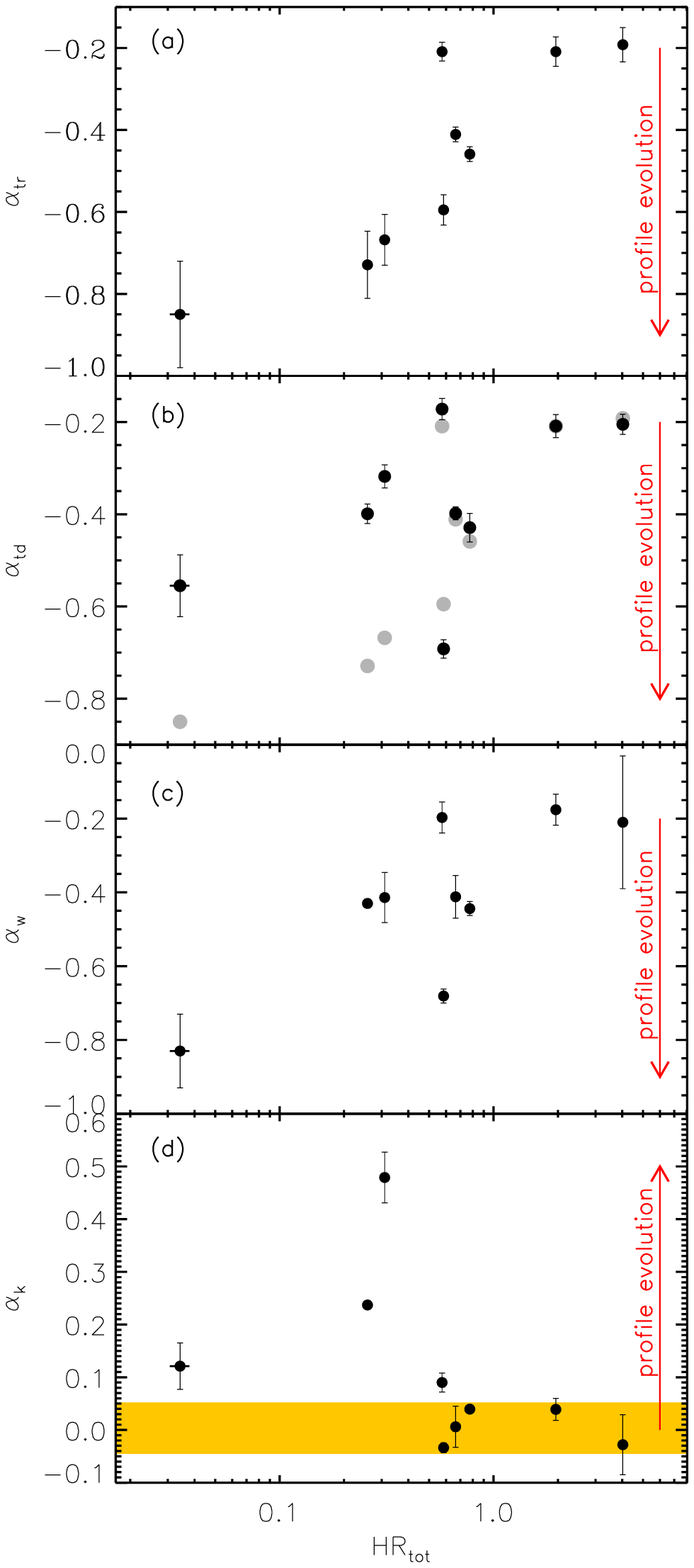}
      \caption{Best fit power-law index describing the evolution of the flare temporal profile parameters
        with observed energy (see Table \ref{Tab:energyevol}) as a function of the spectral hardness of the flare.
        The evolution of each flare parameter with energy band has been fit using the model: $y\propto E^{\alpha}$.
        \emph{(a): }rise time power-law indices. \emph{(b)}: black dots: decay time power-law indices; grey dots:
        rise time power-law indices shown for direct comparison. \emph{(c):} Width power-law indices. \emph{(d):}
      asymmetry power-law indices. The shaded area represents a region of low evolution. In each panel an arrow
      points to the direction of stronger spectral evolution of the parameter.}
\label{Fig:hr_evol_slope}
\end{figure}

\begin{figure}
\vskip -0.0 true cm
\centering
    \includegraphics[scale=0.60]{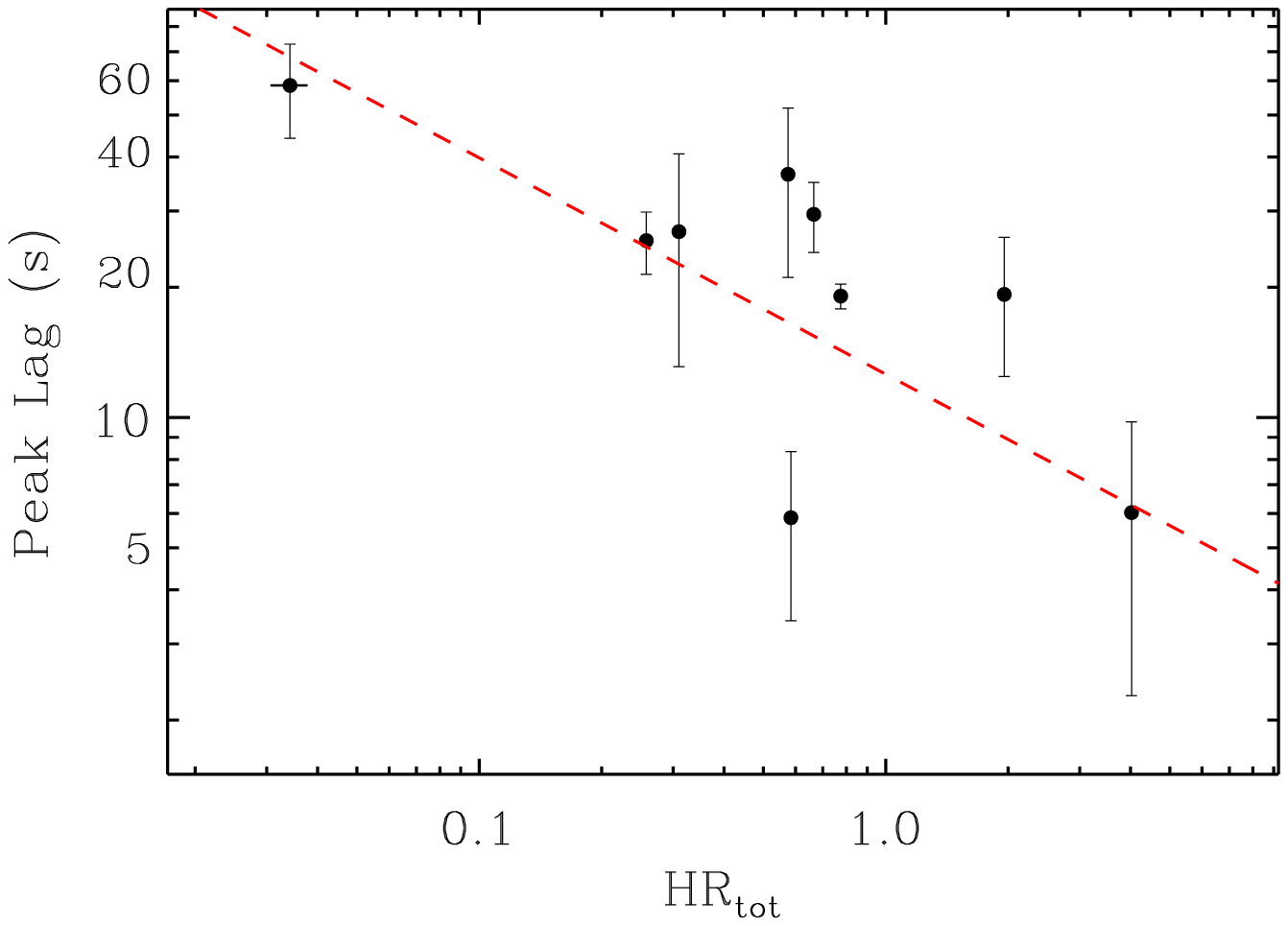}
      \caption{Observed pulse peak lag as a function of the spectral hardness of the flares. A power-law
      with decaying index -0.5 is also shown for comparison (red dashed line).}
\label{Fig:lag_hr}
\end{figure}

The inset of Fig. \ref{Fig:hr_rise_decay} suggests the presence of a correlation between the flare observed 
spectral evolution  and the overall spectral hardness: harder flares
shows a limited level of detected spectral evolution.  In this subsection we prove that a similar conclusion 
is reached through the analysis of the rate of evolution of the temporal parameters describing the flare profiles.

The evolution with energy of each flare profile has been analysed in Subsect. \ref{SubSec:parevolfreq} 
parametrising the evolution of each flare-shape parameter (rise time, decay time, peak time, width 
and asymmetry) using a power-law model. The best fit power-law indices are reported in Table
\ref{Tab:energyevol}  and used as evolution indicators in Fig.\ref{Fig:hr_evol_slope}.
This figure clearly shows that the rate of variation 
of a flare temporal profile is a function of the detected hardness of the emission:
the harder the emission, the lower is the evolution. The parameters describing the profile of softer flares 
undergo a substantial evolution from one energy band to the other (corresponding to higher $|\alpha|$ 
values), while the evolution of hard profiles is limited (lower $|\alpha|$ values). 
Figure \ref{Fig:hr_evol_slope}, panel (b) confirms that the evolution of the rise time 
from one energy band to the other is more sensitive than the decay time to the average spectral 
hardness of the flare. Furthermore, it can be seen that $t_{\rm{r}}$ and $t_{\rm{d}}$ undergo a similar evolution
with bandpass only in hard flares: the softer the flare, the higher is the difference between the 
$t_{\rm{r}}$ and $t_{\rm{d}}$ rate of evolution with energy, with the rise time always evolving faster in energy.
The presence of the trends depicted in the first and second panels from the top of Fig. \ref{Fig:hr_evol_slope}
automatically translates into the results of the third one: since $w=t_{\rm{r}}+t_{\rm{d}}$, it is not surprising that
the width is subject to a higher level of evolution in softer flares. No clear correlation is found between the 
rate of evolution of the asymmetry and the spectral hardness: however,  it is worth noting that the 
asymmetry evolves in the four softer flares, while a limited level of evolution or no evolution is detected in the 
five hardest flares (shaded area of Fig. \ref{Fig:hr_evol_slope}, bottom panel). This is a consequence of the 
results above.

At this point, a lag-hardness correlation is expected:
Fig. \ref{Fig:lag_hr} demonstrates that softer flares are characterised by larger time lags.
\subsection{Spectral modelling}
\label{SubSec:specmodeling}

\begin{figure}
\vskip -0.0 true cm
\centering
    \includegraphics[scale=0.43]{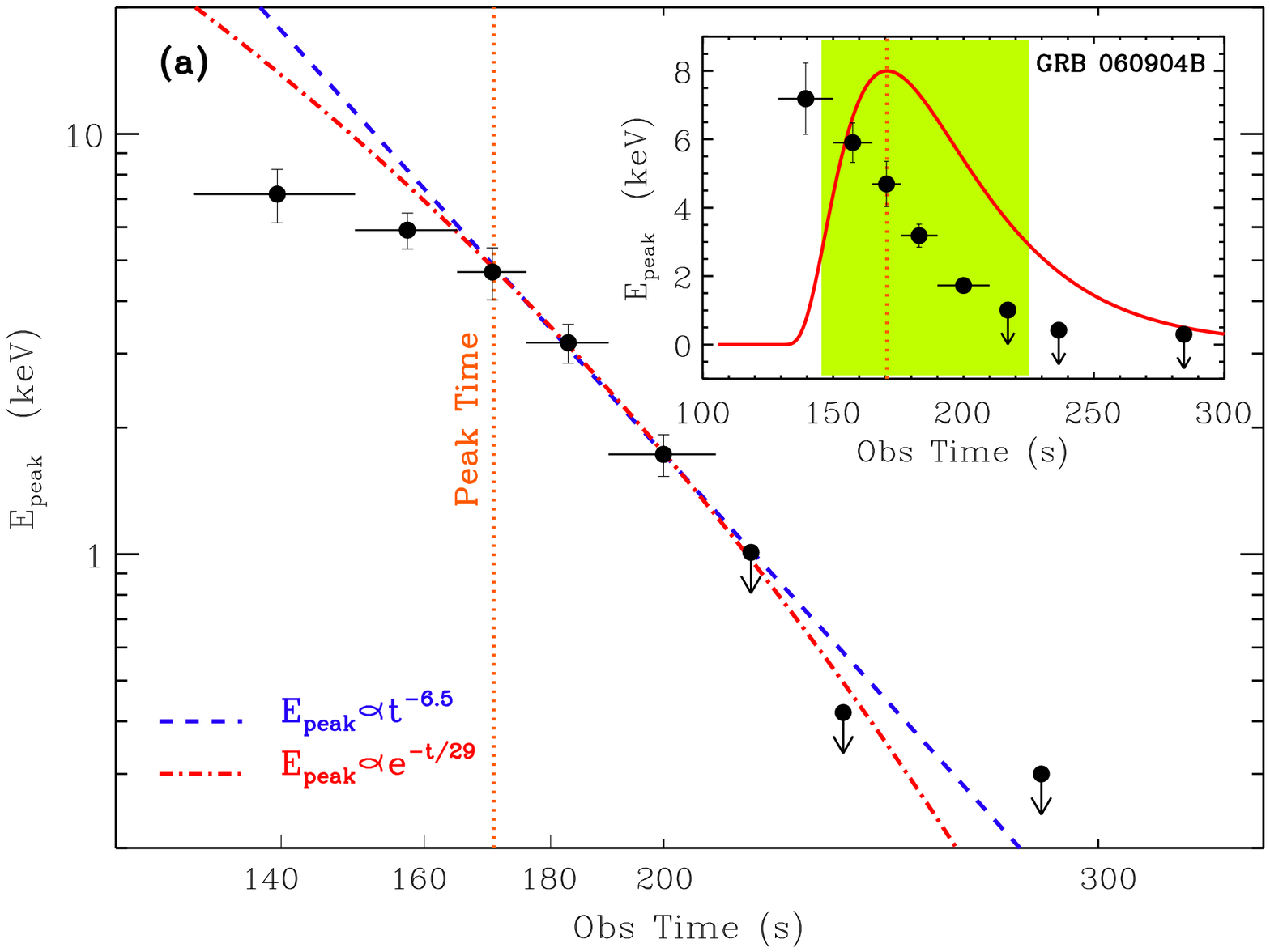}
     \includegraphics[scale=0.43]{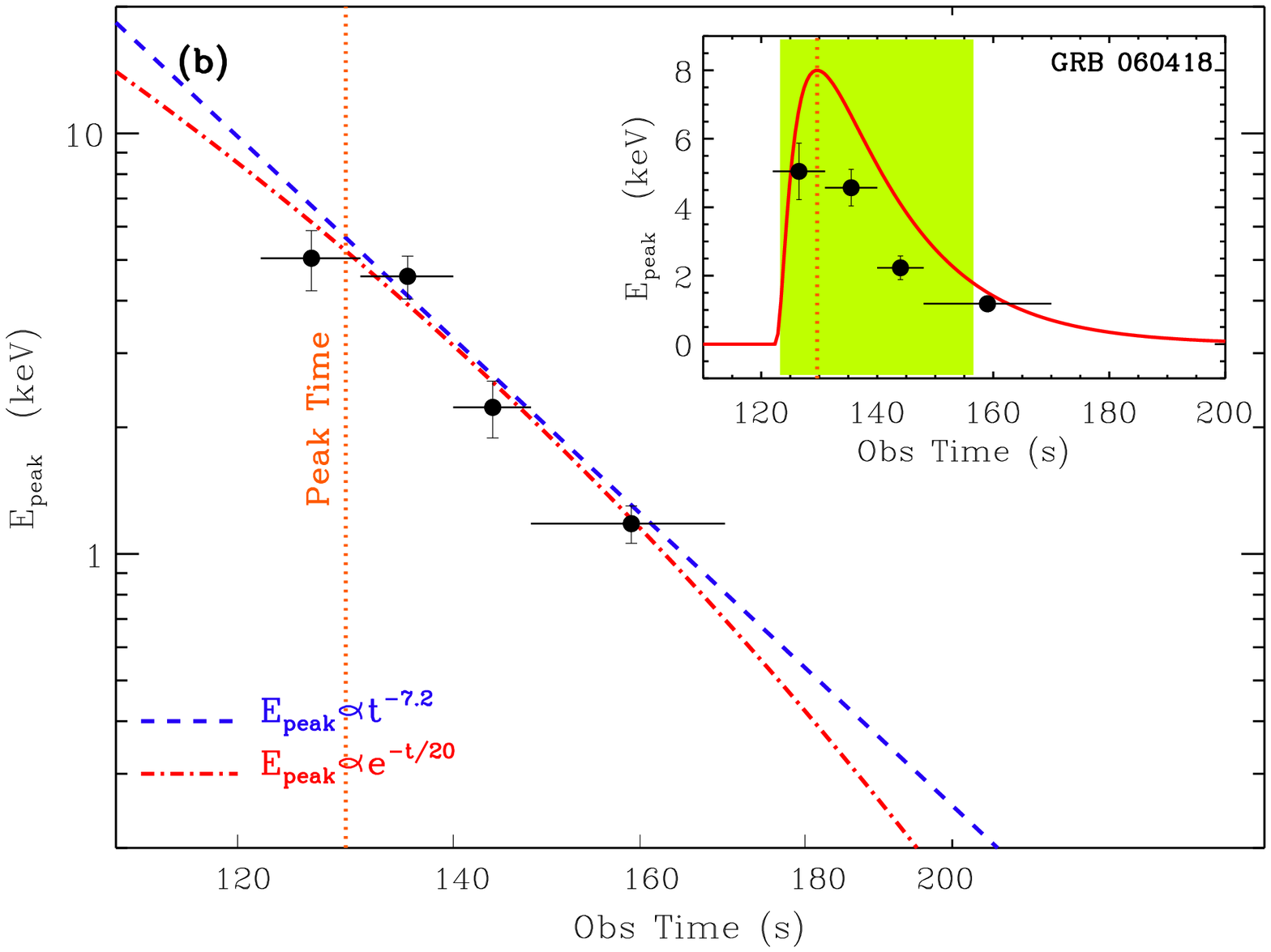}
      \caption{Temporal evolution of the observed peak energy of the $\nu F_{\nu}$ spectrum of the flare detected in
      GRB\,060904B (panel \emph{(a)}) and in GRB\,060418 (panel \emph{(b)}).
       Blue dashed line: best fit power-law decay for the evolution of $E_{\rm{p}}$ during the decay time.
       The best fit power-law decay index is $\alpha_{d}=-7.2\pm0.7$ (GRB\,060418) and $\alpha_{d}=-6.5\pm0.7$ (GRB\,060904B).
       Red dot-dashed line: best fit exponential decay of $E_{\rm{p}}$ during the flare decay time. The zero time of the power-law
	is assumed to be the BAT trigger time. The best fit e-folding time
       is found to be $\tau=20\pm2$ s and $\tau=29\pm 5$ s  for GRB\,060418 and GRB\,060904B, respectively.
      \emph{Insets:} Temporal evolution of $E_{\rm{p}}$ compared to the evolution of the 0.3-10 keV flare profile (red solid line) .
      The flare flux is represented in arbitrary units.  The shaded areas mark the interval of time of extraction of the average 
      flare spectrum, from $t_{\rm{peak}}-t_{\rm{r}}$  to $t_{\rm{peak}}+t_{\rm{d}}$.
      In each panel, the orange dotted lines mark the flare peak time as derived from  
      the best fit parameters.}
\label{Fig:Epeak060904B}
\end{figure}

\begin{table}
  \centering
    \caption{Average spectral properties of the flares detected in GRB\,060418 and GRB\,060904B. The two spectra 
    have been extracted in the time interval defined by ($t_{\rm{peak}}-t_{\rm{r}}$)-($t_{\rm{peak}}+t_{\rm{d}}$) which corresponds to the 
    shaded area in the insets of Fig. \ref{Fig:Epeak060904B}. The fitting model is a photo-electrically absorbed
    Band function with $E_{\rm{p}}$ as free parameter (\textsc{tbabs*ztbabs*ngrbep} within \textsc{xspec}). The intrinsic
   hydrogen column density $N_{\rm{H,z}}$ is frozen to the value found from the joint fit of the time resolved spectra
   (see Table \ref{Tab:spec060904B} and \ref{Tab:spec060418}). The Galactic absorption in the direction of the bursts
   has been accounted for following Kalberla et al., 2005. The isotropic equivalent energy $E_{\rm{iso}}$, 
   the isotropic equivalent luminosity $L_{\rm{iso}}$ and the isotropic equivalent peak luminosity $L_{\rm{p,iso}}$ are computed  in the 
   $1-10^4$ keV rest frame energy range. $L_{\rm{p,iso}}$ is calculated re-scaling the \emph{time integrated} spectrum with the
    flare peak flux. Errors are provided at 90\% c.l.}
    \begin{minipage}{260mm}
      \begin{tabular}{lccccccccccc}
        \hline
             & GRB\,060418& GRB\,060904B\\ 
        \hline
	z & 1.489&0.703 \\
        $\alpha_{\rm{B}}$&   $-0.9$       & $-1.0$ \\ 
        $\beta_{\rm{B}}$ &$-2.30\pm0.09$&  $-2.47\pm0.10$           \\
        $E_{\rm{p}}$ (keV)        &  $3.3_ {-0.3}^{+0.2}$         &   $3.7_{-0.2}^{+0.2}$         \\
        $N_{\rm{H,z}}$ ($10^{22}$ cm$^{-2}$)       &   0.59                         &    0.52                  \\
        $E_{\rm{iso}}$ (erg)   &  $(5.4\pm0.5)\times 10^{51}$           &   $(2.4\pm0.2)\times 10^{51}$        \\
        $L_{\rm{iso}}$ (erg $\rm{s^{-1}}$)   &  $(4.1\pm0.4)\times 10^{50}$           &      $(5.2\pm0.5)\times 10^{49}$   \\
        $L_{\rm{p,iso}}$(erg $\rm{s^{-1}}$) &  $(5.1\pm0.5)\times 10^{50}$  &     $(9.1\pm0.9)\times 10^{49}$     \\
        \hline
        $\chi^2/$dof   & $209.26/216$                &   $312.86/304$        \\
        P-value        & 0.61 &                 0.35     &       \\
        \hline
        \label{Tab:spectot}
    \end{tabular}
    \end{minipage}
\end{table}

The flare emission is typical of the X-ray regime. The limited energy window of the XRT (0.3-10 keV), 
as well as the complete ignorance of the distance of the emitting source coupled to a poor knowledge of the intrinsic 
neutral hydrogen column density absorbing the softest photons, can lead to the misidentification of intrinsically
curved spectra in terms of absorbed power-laws.
We therefore select a sub-sample of flares with measured redshift and with contemporaneous detection by 
the BAT in the 15-150 keV: these requirements restrict the analysis to the two flares observed in GRB\,060904B and
GRB\,060418\footnote{In spite of the measured redshift and BAT detection, GRB\,060526 is excluded
because of the partial overlap of the two structures.}.  

The very good statistics characterising both signals
gives us the possibility to perform a time resolved spectral analysis. To follow the spectral evolution of the
source, we time sliced the BAT and XRT data of GRB\,060418 and GRB\,060904B into 4 and 8 bins, covering 
the 129-320 s and 122-170 s time intervals, respectively. The time intervals were defined so as to contain a
minimum of $\sim$1000 photons in the XRT range and to possibly have enough signal in the BAT energy range to 
constrain the spectral parameters. The spectra were first fitted using an absorbed simple power-law (SPL)
model within \textsc{xspec}: the model is absorbed by both a Galactic and an intrinsic absorption
components. The Galactic component was frozen to the value specified by \cite{Kalberla05} in the direction of the burst
(which is taken to be the UVOT-XRT refined position delivered by the \emph{Swift}-XRT team); the intrinsic 
neutral hydrogen column density $N_{\rm{H,z}}$ was left free to vary. When possible we took advantage of the
simultaneous BAT and XRT coverage, performing a joint BAT-XRT spectral fit. The normalisation  for each 
instrument was always tied to the same value. The results are reported in Table \ref{Tab:spec060418} and Table 
\ref{Tab:spec060904B}. The SPL fit results are statistically unsatisfactory and are not able to account for the 
15-150 keV data. Moreover, the best fit $N_{\rm{H,z}}$ are inconsistent with the values obtained for both GRBs
from late time spectra, where no spectral evolution is detected and no spectral break is expected to lie in the
XRT band pass: for GRB\,060904B and GRB\,060418, from spectra extracted in the time intervals 1-500 ks 
and 200-1000 s, respectively,  we obtained $N_{\rm{H,z}}\sim(0.4\pm0.1)\times 10^{22}\,\rm{cm^{-2}}$ (90\% errors are provided). 
The SPL fits of Table \ref{Tab:spec060418} and \ref{Tab:spec060904B} instead indicate values $>10^{22}\,\rm{cm^{-2}}$. 
These facts, together with the clear evolution of the best fit spectral photon index to softer values with time,
suggest that the peak of the $\nu F_{\nu}$ spectrum is moving through the BAT+XRT bandpass. We tested this possibility 
performing time resolved spectral fits to the empirical Band function (\citealt{Band93}) using $E_{\rm{p}}$ as free parameter 
(\textsc{ngrbep} model). The high and low power-law indices ($\alpha_{\rm{B}}$ and $\beta_{\rm{B}}$) which characterise
the Band model were used as free parameters: we froze $\alpha_{\rm{B}}$ to the typical -1 (see e.g. \citealt{Kaneko06}) value when 
poorly constrained. The uncertainties affecting each parameter were computed freezing $\alpha_{\rm{B}}$ to the best fit value.
The spectra of each flare were fitted simultaneously under the assumption of a constant $N_{\rm{H,z}}$ during the flare
emission. The results of the fitting procedure are listed in Table \ref{Tab:spec060418} and \ref{Tab:spec060904B} for
GRB\,060418 and GRB\,060904B, respectively. The observed evolution of the peak energy with time is represented 
in Fig.\ref{Fig:Epeak060904B}: for both flares, $E_{\rm{p}}$ is found to evolve to lower values, with a decay that 
during the flare decay time can be represented by either a power-law with index $\alpha\sim-7$ or an 
exponential with e-folding time $\tau=29\pm 5$ s (GRB\,060904B) and $\tau=20\pm 2$ s (GRB\,060418).
The uncertainty of the inter-calibration of the BAT and XRT has been investigated as possible source of the detected
spectral evolution: for each time slice, we multiplied the fit model by a constant factor which is frozen to 1 for the
BAT data. For XRT, this factor is left free to vary between 0.9 and 1.1, conservatively allowing the XRT calibration to
agree within 10\% with the BAT calibration. The best fit parameters found in this way are completely consistent with
those listed in Table \ref{Tab:spec060904B} and \ref{Tab:spec060418}. The inter-calibration is therefore unlikely to be the main source of the observed evolution.

\begin{figure}
\vskip -0.0 true cm
\centering
 \includegraphics[scale=0.43]{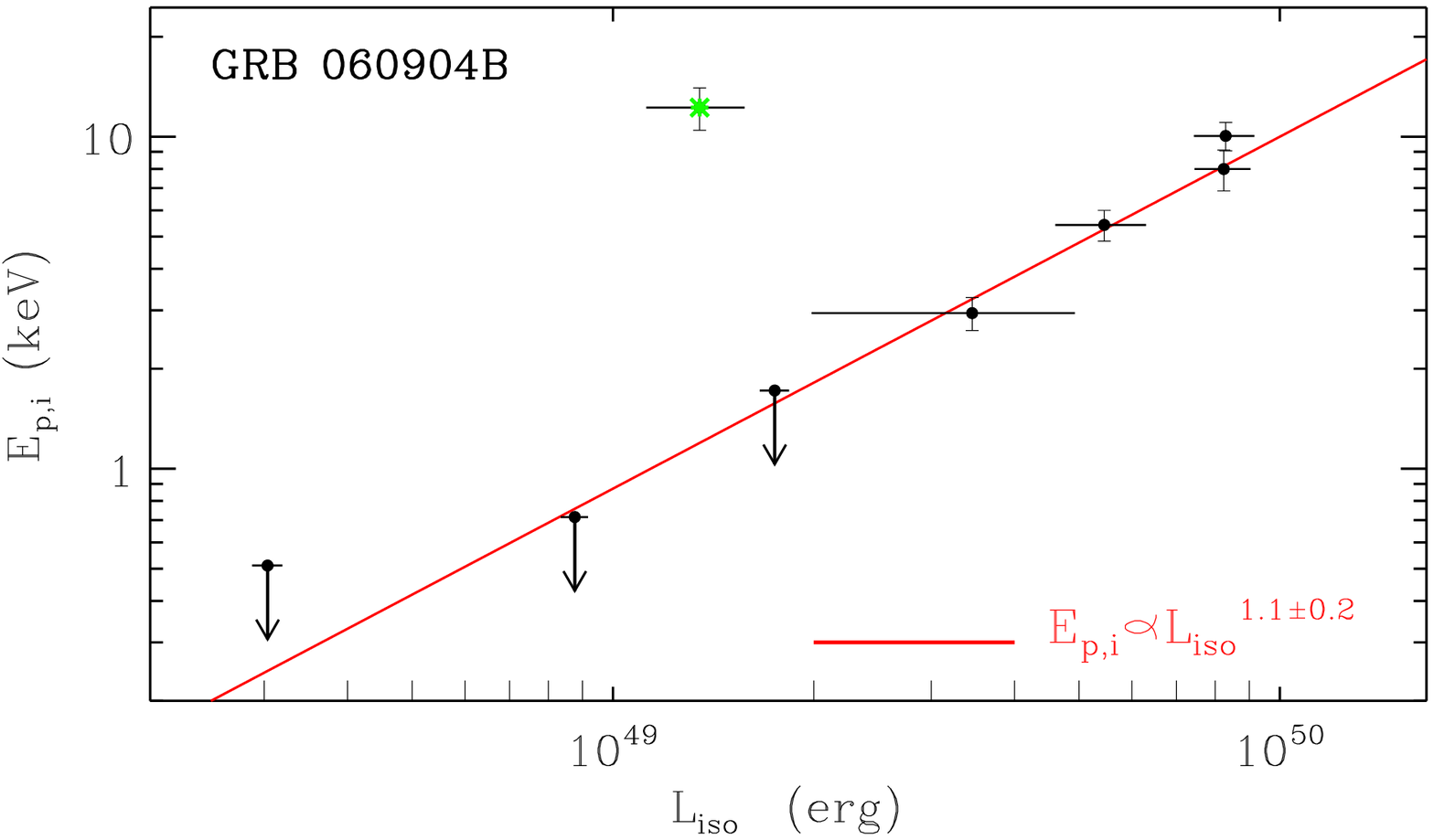}
    \includegraphics[scale=0.43]{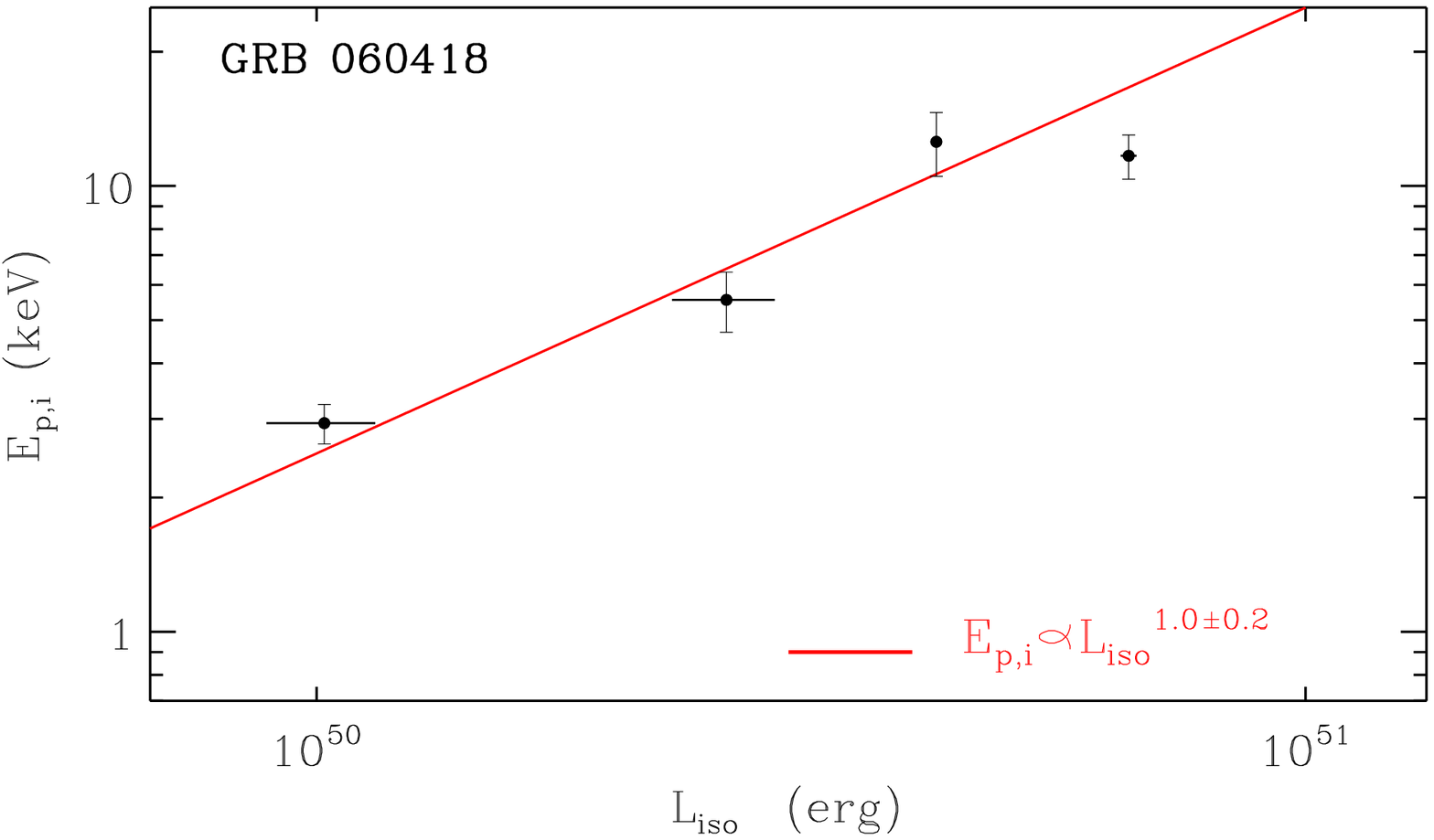}
     \caption{Time resolved $E_{\rm{p,i}}-L_{\rm{iso}}$ correlation for  the flares detected in GRB\,060904B
     (\emph{upper panel}) and  GRB\,060418 (\emph{lower panel}) . Red solid lines: best fit power-law 
     relations: $E_{\rm{p,i}}\propto L_{\rm{iso}}^{1.1\pm0.2}$; $E_{\rm{p,i}}\propto L_{\rm{iso}}^{1.0\pm0.2}$.}
\label{Fig:EpEisoLisotime}
\end{figure}

For each time interval we computed the  isotropic equivalent luminosity $L_{\rm{iso}}$ in the rest frame 1-$10^4$ keV 
energy band: these contributions have been de-absorbed for both the 
Galactic and intrinsic hydrogen column density (see Table \ref{Tab:spec060904B}, \ref{Tab:spec060418}). 
The spectral peak energy is found to evolve  in the $L_{\rm{iso}}-E_{\rm{p,i}}$
plane following a power-law behaviour with best fit power-law index $\sim 1$ (Fig. \ref{Fig:EpEisoLisotime}). 
In particular, for GRB\,060418 (GRB\,060904B) we have $E_{\rm{p,i}}\propto L_{\rm{iso}}^{1.0\pm0.2}$ and 
($E_{\rm{p,i}}\propto L_{\rm{iso}}^{1.1\pm0.2}$)).  For GRB\,060904B we were able to extract a spectrum at the onset of the flare emission:
the properties of this spectrum (green cross in Fig. \ref{Fig:EpEisoLisotime}, upper panel) are clear outliers with 
respect to the general trend defined by the other spectra. The same can not be done for the GRB\,060418 flare: 
in this case, a spectrum extracted at the onset is completely dominated by the contribution of the underlying continuum 
and nothing reliable can be said about the flare spectral properties in this time interval.

The average spectral properties of the two flares have been computed extracting a spectrum from 
$t_{\rm{peak}}-t_{\rm{r}}$ to $t_{\rm{peak}}+t_{\rm{d}}$ (shaded area of the insets of Fig. \ref{Fig:Epeak060904B}): in both cases
the BAT detection rules out a simple power law behaviour ($\chi^2$/dof=289.34/216 P-val=$6\times10^{-4}$ and  
$\chi^2$/dof=466.45/304  P-val=$6\times10^{-9}$, for GRB\,060418 and GRB\,060904B, respectively). 
The two spectra are instead well represented by a 
Band model with best fit observed peak energies $E_{\rm{p}}=3.3_ {-0.3}^{+0.2}$ keV (intrinsic value $E_{\rm{p,i}}=8.2$ keV) 
for GRB\,060418 and $E_{\rm{p}}=3.7_ {-0.2}^{+0.2}$ keV ($E_{\rm{p,i}}=6.3$ keV) for GRB\,060904B (90\% uncertainties are provided). 
The best fit results are reported in Table \ref{Tab:spectot}: note that the isotropic equivalent peak luminosity $L_{\rm{p,iso}}$ is 
calculated starting from the time \emph{averaged} spectrum re-scaled with the flare peak flux. In spite of not 
representing the real flare peak luminosity, this definition allows us to perform a direct comparison to the results obtained
in the literature for the gamma-ray prompt pulses (see e.g. \citealt{Nava08}, \citealt{Yonetoku04}).

\section{Discussion}
\label{Sec:disc}

\begin{figure*}
\vskip -0.0 true cm
\centering
    \includegraphics[scale=0.7]{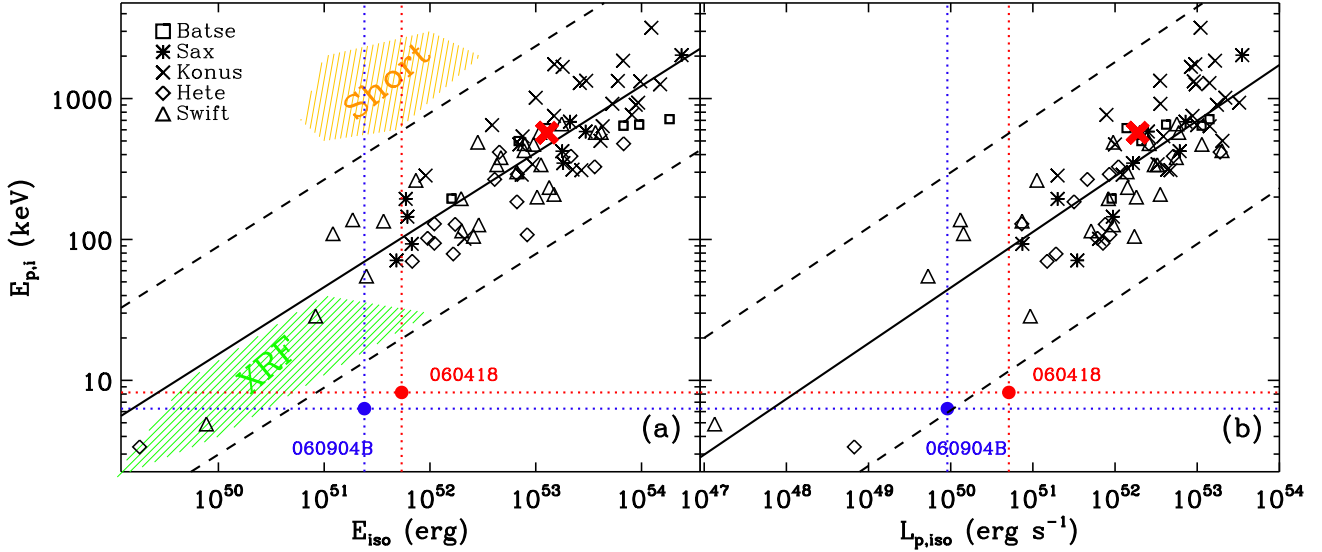}
      \caption{$E_{\rm{p,i}}-E_{\rm{iso}}$ and $E_{p,i}-L_{\rm{p,iso}}$ planes as obtained for the two X-ray flares  detected in
      GRB\,060904B and GRB\,060418 (blue and red dot, respectively) superimposed to the data obtained from the sample
      of 83 GRBs with measured redshift and spectral parameters analysed  in Nava et al, 2008. Black solid lines: best fit to the
      enlarged sample of 110 GRBs published in Ghirlanda et al., 2009, with best fit slopes $\delta=0.476\pm0.025$ and
      $\delta=0.395\pm0.024$ for the left and right panel, respectively. 
      The dashed lines represent the $3\sigma$ scatter.  In both
      panels a red cross marks the \emph{prompt} spectral properties of GRB\,060418. The approximate position of short GRBs and 
      XRFs in the $E_{\rm{p,i}}-E_{\rm{iso}}$ plane is marked with shaded areas.}
\label{Fig:EpEisoLiso}
\end{figure*}

\begin{table*}
  \centering
  \begin{minipage}{180mm}
    \caption{Summary of the relations which link the flare parameters. $\uparrow$ ($\downarrow$) stands for the
    presence of a direct (inverse) proportionality. $\oslash$ means that no correlation has been found
    $L_{\rm{p,iso}}$ is in units of $\rm{erg\,s^{-1}}$; temporal parameters are indicated in s. Errors are reported at 90\% c.l. for correlations involving the peak luminosity.}
  \resizebox{\textwidth}{!}{
    \begin{tabular}{lllllllll}
      \hline
                      & $L_{\rm{p,iso}}$& $t_{\rm{r}}$ & $t_{\rm{d}}$ &$w$ & $k$& $t_{\rm{lag}}$ & HR&$t_{\rm{peak}}$  \\
      \hline
       $L_{\rm{p,iso}}=$&  --            & $10^{(50.9\pm0.4)} t_{\rm{r}}^{((-1.58\pm0.48)}$& $10^{(51.5\pm0.5)} t_{\rm{d}}^{(-1.55\pm0.49)}$ & $10^{(51.8\pm3.9)}w^{(-1.7\pm0.3)}$& $\oslash$ 
     &$10^{(52.82\pm0.20)}t_{\rm{lag}}^{(-0.95\pm0.23)}$&  $\uparrow$& $10^{(54.8\pm4.0)}t_{\rm{peak}}^{(-2.7\pm0.5)}$  \\
       $t_r=$ & --& --& $(2.1\pm0.3)+(0.45\pm0.03)t_d$& $\sim1/3$ & $\oslash$ & $(1.3\pm0.3)+(0.68\pm0.09)t_{\rm{lag}}$ & $ \downarrow$&$(4.0\pm0.3)+(0.06\pm0.01)t_{\rm{peak}}$ \\
       $t_d=$ & --& --& --&$\sim2/3$ & $\oslash$& $(0.2\pm0.7)+(2.2\pm0.3)t_{\rm{lag}}$ &$\downarrow$&$(3.0\pm0.4)+(0.14\pm0.02)t_{\rm{peak}}$\\
       $w=$& --& --& --& --& $\oslash$& $(1.5\pm1.0)+(2.9\pm0.4)t_{\rm{lag}}$  &$\downarrow$&$(4.8\pm0.6)+(0.20\pm0.02)t_{\rm{peak}}$ \\
       $k=$&--& --& --& --& --& $\oslash$&$\oslash$&$\oslash$ \\
       Lag&--& --& --& --& --& --& $ \downarrow$&$ \uparrow$\\
       HR= &  --& --& --& --& --& --& --& $\downarrow$ \\
       \hline
\label{Tab:summary}
  \end{tabular}}
  \end{minipage}
\end{table*}

\begin{table}
	  \centering
  	\begin{minipage}{84mm}
    	\caption{First and second columns: summary of the best fit relations describing the parameters evolution with energy band.
	Third and fourth columns: rate of evolution $|\alpha|$ of the parameters with energy as a function of the spectral hardness: a $\downarrow$
	($\uparrow$) indicates inverse  (direct) correlation. $\oslash$ means that no correlation has been found.}
    	\begin{tabular}{llll}
    		\hline
      &$E$&&HR\\
        \hline
     $w\propto$  &$E^{-0.32\pm0.01}$&$|\alpha_{w}|\propto$& $\downarrow$\\
     $t_{\rm{r}}\propto$&$E^{-0.37\pm0.01}$&$|\alpha_{tr}|\propto$&$\downarrow$\\
     $t_{\rm{d}}\propto$&$E^{-0.29\pm0.01}$&$|\alpha_{td}|\propto$&$\downarrow$\\
     $k\propto$   &$E^{0.11\pm0.01}$&$|\alpha_{k}|\propto$&$\oslash$\\
        \hline
	\label{Tab:summaryenergy}
	\end{tabular}
  	\end{minipage}
\end{table}

Through the analysis of 9 bright X-ray flares we proved the existence of a set of correlations which link both the
temporal and spectral flare parameters. The choice of this sample  introduces an obvious caveat in our analysis,
namely, that we make the explicit assumption that the results we derive from the
bright flares hold for all the flares. However, apart from the brightness, there is no evidence that our sample of bright
flares is not representative of the entire population at the 80\% c.l. as obtained by a Kolmogorov-Smirnov (KS) test.
We therefore conclude that our results are likely to be extended to the entire population of X-ray flares.
\subsection{Temporal parameters evolution with energy}
Flares have different temporal profiles in different energy bandpasses: in strict analogy to prompt emission pulses,
flares are broader and peak later at lower energy. C10 found $w\propto E^{-0.5}$. 
Our sample of bright flares shows $w\propto E^{-0.3}$. For the prompt pulses the relation reads: $w\propto E^{-0.4}$ 
(\citealt{Fenimore95}, but see also  \citealt{Borgonovo07}, N05, \citealt{Norris96} and references therein). The width 
evolution results from the joint evolution of two time scales: the rise and the decay times. The evolution with energy 
of $t_{\rm{r}}$ and $t_{\rm{d}}$ is well described by a power-law, with power-law indices which are found to be very similar 
(see Table \ref{Tab:energyevol}): however, the very good statistics 
of our sample allows us to conclude that the \emph{rise} time is more sensitive to the energy bandpass than the 
decay time. This is evident from Fig. \ref{Fig:trisetdecayslope}. Since the evolution of the temporal parameters with 
energy band is strictly linked to the spectral evolution during the flare emission, this would point to some spectral 
differences between the rise and decay portions of the light-curve.  In particular, the different relative position of the
evolving $E_{\rm{p}}$ with respect to the fixed energy bands during the $t_{\rm{r}}$ and $t_{\rm{d}}$ is likely to play a major role:
this can be in turn a footprint of two different physical mechanisms dominating the rise and the decay phases.
The first consequence of this finding is that each energy channel is only nearly an exact time-stretched version of the others:
flares can be treated as self-similar in energy only at the 0th level of approximation. The second direct consequence 
is the presence of a slight trend pointing to more asymmetric flares at higher energy (Table \ref{Tab:summaryenergy},
Fig. \ref{Fig:par_energy}, panel \emph{d}). \cite{Norris96} defining the asymmetry as $t_{\rm{r}}/t_{\rm{d}}$ found
no evolution of this ratio with energy for GRB prompt pulses: however, the separate evolution  $t_{\rm{r}}(E)$ and $t_{\rm{d}}(E)$ 
was not calculated.

While the soft energy flare profiles are clearly broader and peak later, it is unclear if the flares start simultaneously 
at all energies. The Pulse Start Conjecture was first proposed by \cite{Nemiroff00} and then tested by \cite{Hakkila09}:
these authors studied a sample of 199 prompt BATSE pulses in 75 different GRBs using an automatic pulse fitting methodology.
Each pulse is modelled by a N05 profile (the same profile used for flares).
We caution that a strong coupling is built in the N05 profile between $t_s$ and $\tau_1$:  the two
parameters are clearly related since both fit the flares before the peak. This translates in a non-negligible probability
of obtaining a good fit with perfect $\chi^2$ but with non-physical $t_s$ and $\tau_1$: any slight but systematic variation
$t_s(E)$ would be destroyed. The other 2 major problems are connected to the structure overlap and low statistics
that affect part of the \cite{Hakkila09} sample as discussed by the authors. In spite of these limitations the Pulse Start 
Conjecture was concluded to hold for prompt gamma-ray pulses with an uncertainty of $\sim 0.4$ s. 
For the X-ray flares the situation is different: the selection of 7 isolated flares with very good statistics led to 
the conclusion that a slight trend for flare profiles at \emph{high} energy to start \emph{later} is present.
An example is shown in Fig. \ref{Fig:060904B}, panel \emph{l}. Table \ref{Tab:bestfit} reports the start time values
for the entire sample of bright flares: the high energy profile $t_{s}$ is always larger than the low energy $t_{s}$:
in 5 cases out of 7 the difference between the two values is significant at more than $3\,\sigma$ c.l.

The flare emission preferentially builds up at lower energies at the beginning of the GRB flares.

Figure \ref{Fig:par_energy} demonstrates that the rate of evolution of the temporal parameters with energy in a single
flare can be markedly different from the average sample behaviour (see the values reported in Table \ref{Tab:energyevol}
compared to those of Table \ref{Tab:summaryenergy}). 
Notably, this is directly linked to the observed flare spectral properties as proven by Fig. \ref{Fig:hr_evol_slope}: 
the harder the flare, the lower is the detected rate of evolution of the flare profile in different energy bands.  
Since \emph{hard} flares also
display a \emph{limited} spectral evolution in the 0.3-10 keV energy range (insets of Fig. \ref{Fig:hr_rise_decay}),  
(likely because their $E_{\rm{p}}$ lies outside the XRT energy band during the majority of the emission time), 
the result above implies a strict link between  the spectral properties and the evolution of the temporal
properties. 

The same is true when the flare peak lag is considered: 
in the case of the prompt emission, the cross-correlation lag between energy bands was shown to anti-correlate with the
BATSE spectral hardness ratio by \cite{Norris00}; when \emph{pulses} properties are considered instead of time integrated 
\emph{burst} properties, the result is that the pulse spectral hardness anti-correlates with pulse lag and duration, and
correlates with pulse intensity (\citealt{Hakkila08}).
Figure \ref{Fig:lag_hr} demonstrates that X-ray flares
share the same property: a trend is evident for flares with larger $\rm{HR_{tot}}$ to have
shorter peak lags. The lag-hardness correlation of this figure, together with Fig. \ref{Fig:hr_rise_decay}
again  implies a robust connection between the flares
temporal properties and their spectral evolution: a stronger spectral evolution directly translates 
into a higher lag. This means that the lag characterising the flares profiles is the direct result 
of their spectral evolution. The same conclusion was reached by \cite{Kocevski2003} for the prompt emission:
these authors concluded that the fundamental origin of the observed lag is the evolution of the GRB prompt
spectra to lower energies. Our results of Fig. \ref{Fig:hr_evol_slope} extend this finding to other flare
temporal properties: the rate of evolution with energy of the rise time, decay time and width is inversely 
correlated to the spectral hardness and directly linked to the spectral evolution. 

Finally we mention that only soft flares have been found to be highly asymmetric (Fig. \ref{Fig:par_hr}, 
panel \emph{c}). Similarly \cite{Norris96} found that soft prompt pulses are on average more
asymmetric.

\subsection{The flare lag-luminosity relation}
\label{SubSec:flarelag}
\begin{figure*}
\vskip -0.0 true cm
\centering
    \includegraphics[scale=0.9]{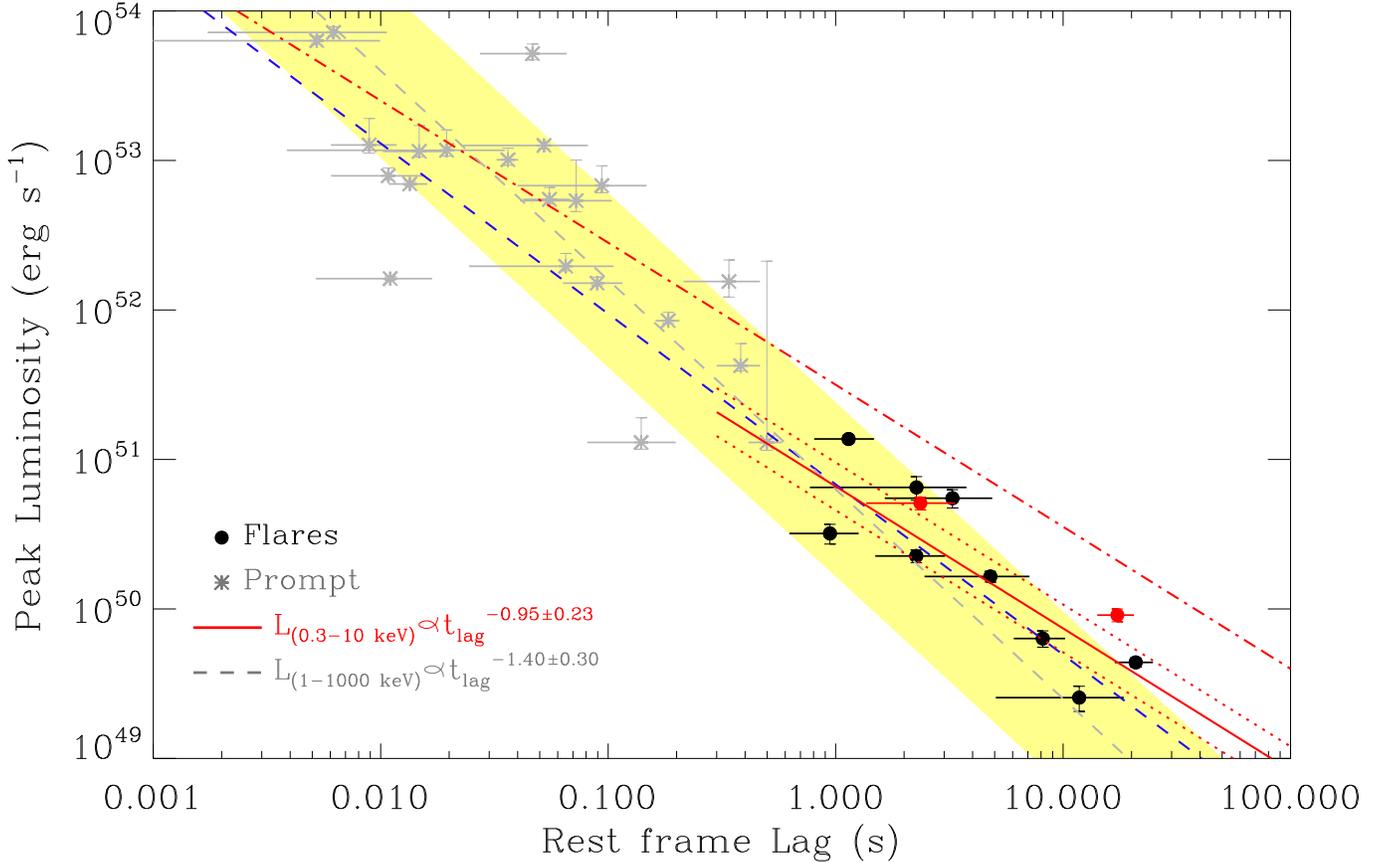}
      \caption{Flare peak-lag peak-luminosity relation. Black filled circles: flares from C10. The isotropic peak luminosity has been computed in the observed 0.3-10 keV
      energy band. 
      The two flares (in GRB\,060418 and GRB\,060904B) for which it was possible to estimate the Band parameters are marked with red bullets: in these
      cases the 1-10000 keV peak luminosity is plotted (see Table \ref{Tab:spectot}). In both cases the flare peak lag is reported.
      Red solid line: flare best fit peak-lag peak-luminosity relation: errors are provided at 90\% c.l. 
      Red dotted lines: best fit sample variance. Grey stars: prompt gamma-ray data corresponding to the
       gold and silver samples of long GRBs from Ukwatta et al., 2010. The isotropic peak luminosity has been calculated in the 1-10000 keV frame. 
       The lag corresponds to the time difference between light-curve structures in the 50-100 keV and 100-200 keV channels. Grey dashed line: 
       prompt best fit peak-lag peak-luminosity relation from Ukwatta et al., 2010. The shaded area marks the 68\% region around this fit. Red dot-dashed line: best fit relation of the
      gold sample performed accounting for the sample variance 
      Blue dashed line: Norris 2000 lag luminosity relation. }
\label{Fig:laglum}
\end{figure*}

\begin{table}
	  \centering
  	\begin{minipage}{84mm}
    	\caption{Prompt $\gamma$-ray lag-luminosity relations.}
  	\resizebox{\textwidth}{!}{
    	\begin{tabular}{lll}
    		\hline
		&Sample &Ref. \\
		\hline
		$\frac{L_{\rm{p,iso}}}{\rm{(erg\,s^{-1})}}=10^{(51.5\pm0.4)}\Big(\frac{t_{\rm{lag},\gamma}}{s}\Big)^{(-0.95\pm0.30)}$ & BAT &1\footnote{
		Gold and silver sample from \cite{Ukwatta10} fitted accounting for the sample variance, 90\% c.l.}\\
		$\frac{L_{\rm{p,iso}}}{\rm{(erg\,s^{-1})}}=10^{(51.54\pm0.05)}\Big(\frac{t_{\rm{lag},\gamma}}{s}\Big)^{(-0.62\pm0.0.04)}$& BATSE& 2\footnote{\cite{Hakkila08}, on single pulses}\\
		$\frac{L_{\rm{p,iso}}}{\rm{(erg\,s^{-1})}}\sim10^{53.1}\Big(\frac{t_{\rm{lag},\gamma}}{0.01\,s}\Big)^{(-1.14\pm0.10)}$ & BATSE&3\footnote{\cite{Norris00}} \\
              \hline
	\label{Tab:laglum}
	\end{tabular}}
  	\end{minipage}
\end{table}

The flare peak lag is directly related to the flare duration (Table \ref{Tab:summary}, Fig. \ref{Fig:Obslag_corr}) and 
since $w$ has been found to vary inversely with the flare peak luminosity (C10, but see also
the relation reported in Table \ref{Tab:summary}), an inverse correlation between peak lag and luminosity is expected.
Since only three flares in our sample have measured redshift, we selected the flares with the best statistics, 
measured redshift and well defined peak times in both the hardest and softest XRT energy bands from the sample of C10. 
Figure \ref{Fig:laglum} confirms the expectation above, showing that the X-ray flares define a rather tight lag-luminosity relation:
\begin{equation}
\frac{L_{\rm{p,iso}}^{\rm{0.3-10\,keV}}}{\rm{(erg\,s^{-1})}}=10^{(50.82\pm0.20)}\times\Big(\frac{t_{\rm{lag,x}}}{\rm{s}}\Big)^{(-0.95\pm0.23)}
\end{equation}
where $t_{\rm{lag,x}}$ is the rest frame peak lag calculated as the difference in time between the flare peak times in the 
0.3-1 keV and the 3-10 keV energy bands. The subscript $x$ reminds that the time lag is calculated in the X-ray regime.
We properly account for the sample variance following the method outlined by \cite{Dagostini05}.
Errors are provided at 90\% c.l.
It is not possible to constrain the spectral peak energy and the spectral slopes of the majority of the flares of C10: for this reason 
we fitted the spectrum extracted around the peak time with an absorbed SPL within \textsc{XSPEC} and conservatively quote 
the isotropic peak luminosity
as obtained in the observed 0.3-10 keV bandpass (black bullets in Fig. \ref{Fig:laglum}). In two cases, for GRB\,060418 and 
GRB\,060904B the Band parameters were determined thanks to the BAT+XRT coverage (Table \ref{Tab:spectot}): this allows us to 
plot the 1-10000 keV (rest frame) isotropic peak luminosity. However, the best fit relation was derived using their 0.3-10 keV
values for homogeneity (the two red bullets in Fig. \ref{Fig:laglum} consistently lie above the expectation). In the cases
of the GRB\,060418 and GRB\,090904B flares, the isotropic peak luminosity in the 0.3-10 keV observed band underestimates the
1-10000 keV rest frame value of a factor 2-3.

The lag-luminosity is one of the key relations which connects the GRB \emph{prompt} temporal and spectral properties: discovered by 
\cite{Norris00} as a time integrated property of each particular burst, the relation was conclusively demonstrated  
to reflect \emph{pulses} rather than \emph{bursts} properties by \cite{Hakkila08}. Figure \ref{Fig:laglum} shows a direct comparison
between the flare and the prompt properties in the lag-luminosity diagram: this is of particular interest since long and short bursts
are known to occupy different regions of the plane (see e.g. \citealt{Gehrels06}). Flares in long GRB are consistent with the long GRBs
lag-luminosity relation. However, we should consider that: first, in Fig. \ref{Fig:laglum} the lag of the prompt data is calculated 
using the Cross Correlation Function (CCF) to the entire BAT light-curve and consequently reflects a time integrated property (see
\citealt{Ukwatta10} for any detail). Second: for the prompt data the lag is defined as the time difference between light-curve structures 
in the 50-100 keV and 100-200 keV channels. For the flares we calculated the \emph{peak} lag between X-ray energy bands. 
In both cases the lag has been computed between band-passes around the event spectral peak energy. Third: the prompt peak luminosity is
calculated in the 1-10000 keV rest frame energy band, while for the flares the peak luminosity is calculated from the 0.3-10 keV
(observed) bandpass (which is expected to be a factor of 2-3 below the 1-10000 keV value).

With these caveats in mind, it is remarkable that the best fit slope of the flare lag-luminosity is consistent with the \cite{Ukwatta10}
results based on BAT data (see Table \ref{Tab:laglum}): 
$L_{\rm{p,iso}}\propto t_{\rm{lag},\gamma}^{(-0.95\pm0.30)}$ (red dot dashed line in Fig. \ref{Fig:laglum}).
It is not surprising that this relation overestimates the flare luminosity which is computed in a narrower energy window by a factor $\sim 5$.
Only a marginal consistency can be quoted with \cite{Hakkila08} who reported an index $\sim0.6$.
Our findings are instead fully consistent with the \cite{Norris00} law (both in normalisation and index), with a power-law 
index 1.14 (blue dashed line in Fig. \ref{Fig:laglum}): these authors reported lags between BATSE energy bands 100-300 kev 
and 25-50 keV. The same is true if we consider the lag-luminosity power-law index by \cite{Schaefer07} who reported a value of 1.01. 
The prompt lag-luminosity relations are summarised in Table \ref{Tab:laglum}.

This result strongly suggests a common physical mechanism producing both the GRB prompt emission  and
the X-ray flare emission hundreds of seconds later. The lag-luminosity relation has been proven to be a fundamental law extending $\sim5$ decades in energy and $\sim5$ decades in time.  
\subsection{Flare parameters correlations}
\label{SubSec:flareparcorr}
At the 0th level of approximation, the flare phenomenology summarised in Table \ref{Tab:summary}  reduces to a set of 
4 independent relations (namely Eq. \ref{Eq:trtd}, \ref{Eq:wtpeak}, \ref{Eq:laglum} and \ref{Eq:softening}). This consistently limits the available 
parameters space for any theoretical model aiming at explaining 
the flare phenomena.  

The first fundamental relation expresses the concept that flares profiles tend to be self-similar both in energy and 
in time (C10):
\begin{equation}
	\label{Eq:trtd}
	t_{\rm{r}}\sim0.5\,t_{\rm{d}}
\end{equation}
which implies: $\tau_1\sim4 \tau_2$: the N05 profile contains a degree of freedom which is not required by the data, since
the rising and decaying time scales are not independent from one another; $t_{\rm{r}}\sim w/3\sim\tau_2$, $t_{\rm{d}}\sim 2w/3\sim\ 2 \tau_2$ and $k\sim const$.
The pulse peak lag between two energy bands centered on $E_1$ and $E_2$ reads: 
$t_{\rm{lag}}(E_1,E_2)=t_s(E_2)+2\tau_{2,E_2} -t_s(E_1)-2\tau_{2,E_1}\sim2(\tau_{2,E_2}-\tau_{2,E_1})$ since $t_s(E_2)\approx t_s(E_1)$ 
at the first level of approximation. $\tau_2$ directly inherits the energy dependence of $t_r$ and $t_d$  (Table \ref{Tab:summaryenergy}): 
$\tau_{2,E_2}=(E_2/E_1)^{-\alpha_\tau}\times \tau_{2,E_1}$ where $\alpha_\tau\sim 0.2-0.7$. This implies $t_{\rm{lag}}\sim f(E_1,E_2)\times\tau_2$
where $f$ is a factor of proportionality  which depends  on $E_1$ and $E_2$.  A lag-width correlation is consequently expected since $w\propto \tau_2$
(or alternatively a $t_{\rm{lag}}-t_{\rm{r}}$ or $t_{\rm{lag}}-t_{\rm{d}}$ relations). 

The width-peak time linear correlation discussed by C10, introduces the concept of temporal $evolution$, which
is the key to interpret the GRB X-ray flare phenomenology:
\begin{equation}
	\label{Eq:wtpeak}
	w\propto t_{\rm{peak}}
\end{equation}
which automatically implies $t_{\rm{r}}\propto t_{\rm{peak}}$ and $t_{\rm{d}}\propto t_{\rm{peak}}$; considering the lag-width correlation of the 
previous paragraph, Eq. \ref{Eq:wtpeak} translates into a lag-tpeak correlation.

The third fundamental relation links the flare peak-lags to their peak-luminosities, in strict analogy to the prompt emission:
 \begin{equation}
	\label{Eq:laglum}
	L_{\rm{p,iso}}\propto t_{\rm{lag}}^{-0.95\pm0.23}
\end{equation}
Consequently, $t_r$, $t_d$ and $w$ will be linked to the flare peak luminosity. In turn, the peak luminosity is expected to evolve with time to lower values,
as found: $L_{\rm{p,iso}}\propto t_{\rm{peak}}^{-2.7}$ although with a large scatter.

The fourth relation again describes the temporal \emph{evolution} of the flare spectral properties:
 \begin{equation}
	\label{Eq:softening}
	\rm{HR}(t)\propto t_{\rm{peak}}^{-\alpha}
\end{equation}
Flares become softer and softer as time proceeds. As a result, an inverse $\rm{HR}-t_{\rm{r}}$, $\rm{HR}-t_{\rm{d}}$, $\rm{HR}-w$ and $\rm{HR}-t_{\rm{lag}}$ relation is 
automatically built. This completes the set of 28 correlations (or lack thereof) of Table \ref{Tab:summary}.  
An important implication is that at the 0th level of approximation,
the physics underlying the flare emission determines that out of 3 time scales describing the flare temporal profile 
(namely $t_{\rm{r}}$, $t_{\rm{d}}$ and $t_{\rm{peak}}$ ), only one degree of freedom survives. 

As a second level approximation, one should consider that $t_s=t_s(E)$; $f=f(E_1,E_2,\alpha_\tau)$ which likely reflects the relative position of the
spectral peak energy with respect to  $E_1$ and $E_2$; $k=k(E)$ since the rise and decay times evolution with energy band slightly 
differs from one another(Fig. \ref{Fig:trisetdecayslope}). Any theoretical model aiming at explaining the flare emission is asked to be consistent
with these set of findings, together with the flare spectral properties discussed in the following paragraph. A critical revision of 
the existing models is in preparation.

\subsection{Temporal evolution of a flare spectrum}
\subsubsection{$E_{\rm{p}}(t)$ }
For the two flares with the highest statistics, we find that the flare spectrum is well described by an 
evolving Band function.
In both cases the spectral peak energy evolves with time to lower values following an exponential decay which 
tracks the decay of the flare flux. After $t_{\rm{peak}}$ the N05 profile is progressively dominated by the 
$\rm{exp}(-t/\tau_{2})$ factor, with the relative strength of the $\rm{exp}(-\tau_{1}/t)$ dropping
quickly. Remarkably, in both flares, the temporal evolution of the spectral peak energy and the flux
seem to share the same e-folding time: for the flare detected in GRB\,060418 we have $\tau_2=22.2\pm0.4$ s with 
$\tau_{Ep}=20.0\pm2.0$ s; for the GRB\,060904B flare we find $\tau_2=29.5\pm0.6$ s in good agreement with 
$\tau_{Ep}=29.0\pm4.7$ s. 

The evolution of $E_{\rm{p}}(t)$ to lower values during the pulse decay time is one of the signatures of the prompt
emission as demonstrated by a number of studies (see e.g. \citealt{Peng09} and references therein for a recent 
study on single prompt pulses). The spectral peak energy of prompt pulses is found to evolve following a 
power-law\footnote{The zero-time is chosen to be the starting point of the rising segment.} 
$E_{\rm{p}}(t)\sim t^{-\delta}$ with an evolutionary slope $\delta\sim 1$ (see \citealt{Peng09} and references therein). 
The observed evolution of the flare $E_{\rm{p}}(t)$ is instead much faster and inconsistent with the $t^{-1}$ behaviour 
after rescaling the time to the beginning of the flare emission. Taking the GRB\,060904B flare which starts at 
$t_s\sim124$ s as an example, $E_{\rm{p}}(171\,\rm{s})\sim 4.7$ keV  would imply $E_{\rm{p}}(300\,\rm{s})\sim 1.2$ keV 
if the $t^{-1}$ behaviour  is assumed: the observations instead constrain the $E_{\rm{p}}(300\,\rm{s})$ to be well
below the XRT bandpass, so that $E_{\rm{p}}(300\,\rm{s})<0.3$ keV. 
While this could be the result of a wrong choice 
of the power-law zero time (the relevant time could be the ejection time instead of $t_{\rm{s}}$, see \citealt{Willingale10}
for details), further investigations 
are required to establish if the faster $E_{\rm{p}}(t)$ evolution
is a peculiar feature of the flare emission when compared to the prompt phenomenology.
\subsubsection{Flares in the $E_{p}-E_{\rm{iso}}$ and $E_{p}-L_{\rm{p,iso}}$ planes}

The spectral analysis of Subsec. \ref{SubSec:specmodeling} allows us to constrain the properties of two X-ray flares
in the $E_{\rm{p}}-E_{\rm{iso}}$ and $E_{\rm{p}}-L_{\rm{p,iso}}$ planes for the first time. Figure \ref{Fig:EpEisoLiso} shows
the result: when compared to the spectral properties of the prompt emission of 83 GRBs with measured redshift
and well constrained spectra (\citealt{Nava08}; \citealt{Ghirlanda09}), the two flares show  higher 
$E_{\rm{iso}}$ and $L_{\rm{p,iso}}$ than expected starting from their rest frame spectral peak energy $E_{\rm{p,i}}$.  
If the flares are part of the prompt emission, then they are not expected to share the time integrated  $E_{p,i}-E_{\rm{iso}}$ 
best fit normalisation:  however, \cite{Krimm09} demonstrated that individual sequences of the same burst do follow the  
$E_{p}-E_{\rm{iso}}$  relation with a \emph{higher} normalisation (reflecting the fact that the total energy budget is
distributed over the sequence, see \citealt{Krimm09}, their Fig. 14). The opposite is observed for the two flares.
The $E_{\rm{p}}-L_{\rm{p,iso}}$ plane does not suffer from this effect, and time integrated properties can be here directly
compared to their time resolved counterparts. In this plane the two flares are barely inside the $3\sigma$ best fit area: 
however, the peak luminosity is poorly sampled below $5\times 10^{50}\,\rm{erg\,s^{-1}}$ and a deviation to lower
spectral peak energies of the main sample cannot be excluded since the best fit relation is mainly established by high
luminosity data points. In this case the X-ray flares would belong to the same $E_{\rm{p}}-L_{\rm{p,iso}}$ as the prompt emission.
Alternatively, and equally interesting, the flares do not follow the $E_{\rm{p,i}}-L_{\rm{p,iso}}$  relation of the prompt data.
This possibility is suggested by the fact that the two flares are outliers of the  $E_{\rm{p,i}}-E_{\rm{p,iso}}$ relation even when
the effect of the time resolved spectral analysis is taken into account.
Whether this depends on the chosen integration energy band (1-10$^4$ keV) is currently under investigation.

For GRB\,060418 the spectral properties during the prompt emission were measured (see \citealt{Nava08}; \citealt{Amati08}):
$E_{\rm{p,i}}^{\gamma\rm{-ray}}=572\pm143$  keV; $E_{\rm{iso}}^{\gamma\rm{-ray}}=(13\pm3)\times 10^{52}$ erg; 
$L_{\rm{p,iso}}^{\gamma\rm{-ray}}=1.9\times 10^{52}$ erg. The flare isotropic input comprises $\sim 5$\% of the isotropic
gamma ray prompt energy while the flare isotropic peak luminosity is $\sim 3$\% the $L_{\rm{p,iso}}^{\gamma\rm{-ray}}$.
The contribution of the flare emission is therefore well within the uncertainties estimated for the prompt parameters:
this results helps to  understand how much the unaccounted flare emission contributes to the scatter affecting the
$E_{\rm{p,i}}-E_{\rm{iso}}$ relation (\citealt{Amati08}).
\subsubsection{The flare $E_{p}(t)-L_{\rm{p,iso}}(t)$ relation}

The time resolved analysis of the two flares with redshift and BAT detection reveals that the rest frame spectral peak energy
$E_{\rm{p,i}}$ correlates with the isotropic luminosity $L_{\rm{iso}}$ within single
flares. The same has been demonstrated to be true for the prompt pulses: the \emph{time integrated} correlations 
are the result of the existence of similar \emph{time resolved} correlations of the same parameters. 
Individual GRB prompt pulses are consistent with the $E_{\rm{p,i}}-L_{\rm{iso}}$
as proved by \cite{Ghirlanda09}, and  \cite{Ohno09}. Our work proves that 2 X-ray flares share this property and that 
flares in general are likely to be consistent with this behaviour. The best fit slopes of the two relations of Fig.
\ref{Fig:EpEisoLisotime} are however steeper than  the $\sim0.5$ slope found for the
prompt emission (see e.g. \citealt{Ghirlanda09}): whether this is something peculiar of the flare emission or not, needs to be understood
with a larger sample. At the moment we note that the best fit time resolved $E_{\rm{p,i}}-L_{\rm{iso}}$ 
slope within single prompt pulses have been found to be different from the 0.5 value in some cases: see e.g. GRB\,090323 and 
GRB\,090328 in \cite{Ghirlanda09}  their Fig. 3, where the  $E_{\rm{p,i}}-L_{\rm{iso}}$ track seems to be steeper than $0.5$.

Intriguingly, the spectrum extracted at the onset of the flare in GRB\,060904B is a clear outlier,
being characterised by an  $E_{\rm{p,i}}$ higher than expected: a similar result was found by \cite{Ohno09} for the prompt emission. 
These authors  performed a time resolved spectral analysis of the GRB\,061007 prompt emission and concluded that the 
\emph{initial} rising phase of each pulse is an outlier of the $E_{\rm{p,i}}-L_{\rm{iso}}$ relation, with $E_{\rm{p,i}}$ around twice 
the value expected from the spectral correlation. However, no difference between the rise and decay portions of the prompt pulses
has been reported by \cite{Ghirlanda09} from the analysis of 2 \emph{Fermi} GRBs.

The presence of  these spectral correlations during the flare emission tightly links the X-ray flare emission to the prompt phase;
as a by-product, these results strengthen the interpretation of the spectral energy correlations as manifestation of the physics
of the GRBs (see \citealt{Nava08} and references therein for a detailed discussion of this topic).
\subsection{Flares and light-curve morphology}
\begin{figure}
\vskip -0.0 true cm
\centering
    \includegraphics[scale=0.42]{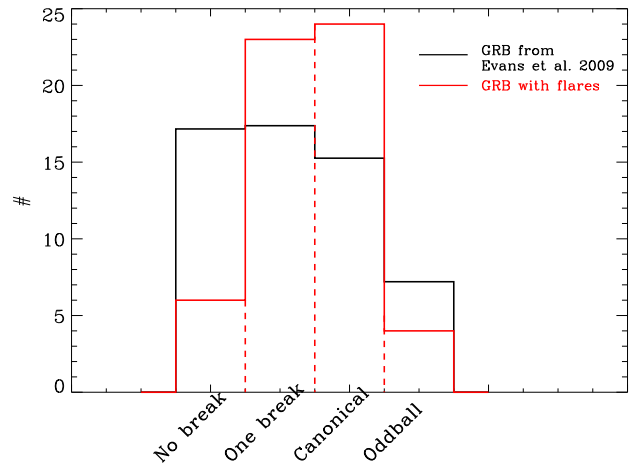}
      \caption{Red line: classification of the GRBs belonging to the flare sample of C10
      according to their  X-ray afterglow morphology  (see Evans et al., 2009  for details). Black line: GRBs belonging to the
      Evans et al. 2009 sample detected in the same period of time. This distribution has been re-normalised for the total number
      of GRBs of the flare sample.}
\label{Fig:flares_lctype}
\end{figure}

The entire set of findings presented in this work establishes a strong link between early X-ray flares and prompt pulses.
Flares are consistent to be independent episodes of emission superimposed to the contemporary steep decay or afterglow 
components: while the first has been proven to be generated by the tail of the previous pulses (\citealt{Kumar00}; 
see e.g. \citealt{Willingale10}) and is therefore related to the physical mechanism which gives rise to the prompt emission, 
the latter is likely to be a completely independent component (e.g. \citealt{Margutti10}).  

X-ray afterglows with no breaks are likely to be dominated by the afterglow component from the beginning 
 of the XRT observations (see \citealt{Liang09} for a dedicated study).  
We therefore expect the SPL  X-ray afterglows to show a limited flaring component if flares really constitute
an independent component (flares would be perhaps present 
but hidden by the contemporaneous afterglow emission). Figure \ref{Fig:flares_lctype} confirms this expectation: 
the no-break class is clearly under-represented in the flare sample, while the vast majority of the detected early time flares
reside in one-break or canonical X-ray afterglows. The probability that  the two samples are drawn from the same population
is evaluated to be as low as $0.9$\% using a KS test. If this is due to the brightness of the afterglow emission, than the conclusion 
is that flares have an independent origin:  in the standard scenario this means that flares are not produced by the external shock. 
Another equally interesting possibility is that flaring emission is quenched (instead of being hidden) in SPL-afterglow GRBs.
Distinguishing between these two possibilities is beyond the scope of this work and  will be the subject of a forthcoming investigation.
\section{Summary and conclusions}
\label{Sec:conc}
\begin{figure}
\vskip -0.0 true cm
\centering
    \includegraphics[scale=0.43]{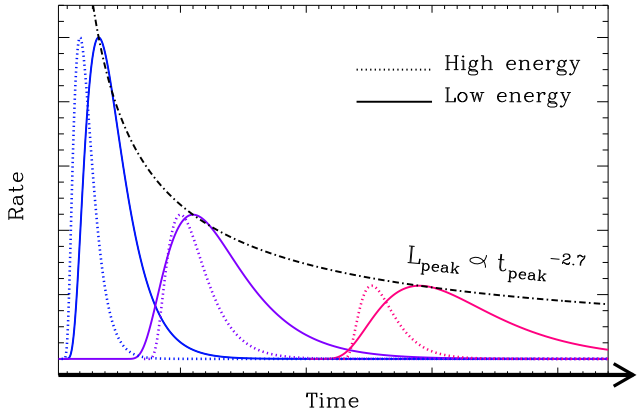}
      \caption{Flare paradigm. Dashed curves: high energy profile. Solid line: low energy profile. High energy profiles
      rise faster, decay faster and peak before the  low energy emission. As time proceeds, flares becomes wider, with higher peak lag,  lower peak
      luminosities and softer emission. However, at the 0th level of approximation 
      the self-similarity is preserved both in energy and in time (with $t_r/t_d\sim0.5$). We stress that this is an \emph{average} behaviour.
      The thick solid arrow underlines the dependence of the flare properties on the \emph{time}, in strong opposition to prompt pulses. }
\label{Fig:flareparadigma}
\end{figure}

The GRB X-ray flare phenomenology shows a list of properties strictly analogous to the gamma-ray prompt emission
(see \citealt{Hakkila08} for a recent study on individual prompt pulses). However, it differs in 
one crucial aspect: flares \emph{evolve} with \emph{time}\footnote{We parenthetically note that this difference 
could be due to the fact that the relevant time for a peak (prompt pulse or flare) is the time since the shell ejection
in a standard internal shock scenario, and not the time since trigger. While these two time scales are completely
unrelated for the prompt pulses, they are more closely related for flares (since the time is large the difference
between the two becomes negligible). See \cite{Willingale10} and Willingale et al., in prep. for further details.}. 
Flares evolve with time to lower peak intensities, larger widths,
larger lags and softer emission. Fig. \ref{Fig:flareparadigma} illustrates the flare paradigm. 
As the prompt emission pulses, flares have correlated 
properties: short-lag flares have shorter duration, are more luminous and harder than long-lag flares. 
In particular the following properties add  to the lists of Sec. \ref{Sec:Introduction}:
\begin{itemize}
\item Flares define a lag-luminosity relation: \\$L_{\rm{p,iso}}^{\rm{0.3-10\,keV}}\propto t_{\rm{lag}}^{-0.95\pm0.23}$. The best 
  fit slope is remarkably consistent with the prompt findings  (see \citealt{Ukwatta10} and references therein for a recent study);
\item The lag is found to correlate with the flares width while it is inversely correlated to the flare spectral hardness;
\item The flares temporal profiles in different band-passes are only a nearly exact time stretched version of one another: the rise and 
  decay times evolve following slightly different power-laws in energy: $t_r\propto E^{-\alpha_{tr}}$, $t_d\propto E^{-\alpha_{td}}$ with 
  $\alpha_{tr}>\alpha_{td}$. The result is that flares are on average more asymmetric at high energy;
\item The rate of evolution of a flare profile in different band-passes anti-correlates with its spectral hardness: the harder the flare
  the lower is the rate of evolution from one energy band to the other;
 \item The flare spectral peak energy $E_{p}(t)$ evolves to lower values following an exponential decay which tracks the decay of the
   flare flux. The detected evolution is faster and inconsistent with the $\sim t^{-1}$ behaviour even when the zero time is re-set to the 
   beginning of the flare emission;
\item The two flares with best statistics show higher than expected $E_{\rm{iso}}$ and $L_{\rm{p,iso}}$ values 
  when compared to  the $E_{\rm{p,i}}-E_{\rm{iso}}$ and $E_{\rm{p,i}}-L_{\rm{iso}}$ prompt correlations; 
\item The rest frame spectral peak energy $E_{\rm{p,i}}$ correlates with the isotropic luminosity $L_{\rm{iso}}$ 
  within single flares, giving rise to a time resolved  $E_{\rm{p,i}}-L_{\rm{iso}}$ correlation;
\item The flare emission preferentially builds up at lower energies: flares do not seem to be consistent with the Pulse Start Conjecture;
\item Among the different types of X-ray afterglow light-curves, the simple power-law afterglows are under-represented in the flare sample. 
  Flares are preferentially detected superimposed to one-break or canonical light-curves.
\end{itemize}
The strict analogy between the prompt pulses and flares phenomenology strongly suggests a common origin of the two phenomena.




\section*{Acknowledgments}
RM is grateful to L. Nava for providing the comparison data
portrayed in Fig. \ref{Fig:EpEisoLiso}. This work is supported by ASI grant SWIFT I/011/07/0,
by the Ministry of University and Research of Italy (PRIN MIUR 2007TNYZXL), by MAE and by the 
University of Milano Bicocca, Italy.

\appendix
\section{Tables}
\setcounter{table}{0}
\begin{table*}
 \centering
  \begin{minipage}{180mm}
  \caption{Flares best fit parameters according to Eq. \ref{Eq:norris05} and derived quantities (Eq. \ref{Eq:tpeak}, 
  \ref{Eq:width}, \ref{Eq:asymmetry} and \ref{Eq:trisetdecay}). The reported uncertainties are calculated accounting
  for the entire covariance matrix for each fit. From left to right: GRB name, energy band of investigation, 
  redshift, normalisation, start time, first shape parameter, second shape parameter, peak time, width, asymmetry, 
  rise time, decay time, chi squared, degrees of freedom. In each column, a -1 indicates the absence of the value in 
  the literature (for the redshift) or the impossibility to constrain the value from the fit.  A negative error indicates that 
  the parameter was first left free to vary in the fit and then frozen to the reported value to constrain the errors 
  associated to the other fit parameters.  }
  \resizebox{\textwidth}{!}{
  \begin{tabular}{llllllllllllll}
  \hline
   GRB & band & z & norm & $t_s$ &$\tau_1$ & $\tau_2$& $t_{\rm{peak}}$& $w$&$k$&$t_r$&$t_d$&$\chi^2$&$dof$\\
           &         &    &($\rm{cts\,s^{-1}})$ & (s)&(s)&(s)&(s)&(s)  & & (s)& (s)   \\    
  \hline
050822 &tot\footnote{0.3-10 keV} & -1.0 &  45.5 $\pm$   1.3 & 389.2 $\pm$   7.5 &  48.0 $\pm$  18.0 &  54.6 $\pm$   3.1 & 440.4 $\pm$   8.3 & 119.0 $\pm$   4.4 & 0.459 $\pm$ 0.007 &  32.2 $\pm$   3.4 &  86.8 $\pm$   2.3 & 283.6 &267 \\
 &       1\footnote{0.3-1 keV}  & -1.0 &  22.3 $\pm$   0.8 & 374.0 $\pm$  13.0 & 123.0 $\pm$  47.0 &  53.6 $\pm$   4.6 & 455.2 $\pm$  12.6 & 142.4 $\pm$  13.2 & 0.376 $\pm$ 0.013 &  44.4 $\pm$   3.8 &  98.0 $\pm$   3.1 & 324.2 &290 \\
 &       2\footnote{1-2 keV}  & -1.0 &  20.2 $\pm$   1.0 & 371.0 $\pm$  16.0 &  85.0 $\pm$  56.0 &  38.3 $\pm$   4.1 & 428.1 $\pm$  16.2 & 101.0 $\pm$  18.0 & 0.379 $\pm$ 0.014 &  31.4 $\pm$   5.3 &  69.7 $\pm$   3.2 & 259.4 &181 \\
 &       3\footnote{2-3 keV}  & -1.0 &   5.0 $\pm$   -1.0 & 375.0 $\pm$   -1.0 &  21.0 $\pm$  19.0 &  22.4 $\pm$   6.8 & 396.7 $\pm$   6.9 &  49.4 $\pm$   4.3 & 0.453 $\pm$ 0.021 &  13.5 $\pm$   2.1 &  35.9 $\pm$   5.2 &  12.7 &18 \\
 &       4\footnote{3-10 keV}  & -1.0 &   -1.0  &  -1.0 &   -1.0 &   -1.0&  -1.0 &   -1.0&  -1.0 &   -1.0 &   -1.0 &   -1.0 & -1.0 \\
060418 &tot &1.489 & 379.1 $\pm$   7.9 & 121.9 $\pm$   0.2 &   2.0 $\pm$   0.3 &  22.2 $\pm$   0.4 & 128.6 $\pm$   0.4 &  32.9 $\pm$   0.4 & 0.673 $\pm$ 0.008 &   5.4 $\pm$   0.3 &  27.5 $\pm$   0.3 &1665.1 &975 \\
 &       1 &1.489 &  52.6 $\pm$   1.6 & 111.8 $\pm$   1.4 &  15.9 $\pm$   4.1 &  31.4 $\pm$   1.2 & 134.1 $\pm$   2.5 &  61.6 $\pm$   1.4 & 0.510 $\pm$ 0.008 &  15.1 $\pm$   1.1 &  46.5 $\pm$   1.1 & 644.4 &497 \\
 &       2 &1.489 & 166.7 $\pm$   4.1 & 118.6 $\pm$   0.4 &   8.3 $\pm$   1.3 &  19.4 $\pm$   0.4 & 131.3 $\pm$   0.9 &  36.9 $\pm$   0.6 & 0.526 $\pm$ 0.005 &   8.8 $\pm$   0.4 &  28.2 $\pm$   0.5 &1328.2 &876 \\
 &       1 &1.489 &  83.8 $\pm$   3.1 & 119.9 $\pm$   0.6 &   8.0 $\pm$   2.1 &  10.9 $\pm$   0.6 & 129.3 $\pm$   1.0 &  22.9 $\pm$   0.6 & 0.475 $\pm$ 0.005 &   6.0 $\pm$   0.4 &  16.9 $\pm$   0.5 & 408.5 &296 \\
 &       4 &1.489 & 121.8 $\pm$   4.9 & 121.1 $\pm$   0.3 &   5.5 $\pm$   0.6 &   9.5 $\pm$   0.4 & 128.3 $\pm$   0.3 &  19.0 $\pm$   0.6 & 0.498 $\pm$ 0.005 &   4.8 $\pm$   0.1 &  14.3 $\pm$   0.5 & 491.0 &308 \\
{060526f\footnote{First flare detected by the XRT.}} &    tot &3.221 & 282.9 $\pm$   5.6 & 225.0 $\pm$   1.9 &  62.0 $\pm$  15.0 &  14.4 $\pm$   1.3 & 254.9 $\pm$   2.3 &  43.9 $\pm$   3.8 & 0.328 $\pm$ 0.006 &  14.8 $\pm$   0.4 &  29.2 $\pm$   1.1 &1024.5 &948 \\
 &       1 &3.221 &  49.7 $\pm$   1.5 & 220.0 $\pm$   0.0 &  95.0 $\pm$   9.3 &  24.5 $\pm$   3.2 & 268.2 $\pm$   1.2 &  73.0 $\pm$   3.8 & 0.336 $\pm$ 0.006 &  24.2 $\pm$   1.4 &  48.7 $\pm$   4.6 & 823.0 &642 \\
 &       2 &3.221 & 143.8 $\pm$   3.4 & 228.0 $\pm$   0.0 &  53.5 $\pm$   2.3 &  15.1 $\pm$   0.7 & 256.3 $\pm$   0.3 &  43.9 $\pm$   0.8 & 0.342 $\pm$ 0.002 &  14.4 $\pm$   0.3 &  29.5 $\pm$   0.9 &1306.2 &850 \\
 &       3 &3.221 &  64.2 $\pm$   2.2 & 230.0 $\pm$   0.0 &  35.6 $\pm$   2.3 &  12.7 $\pm$   0.6 & 253.3 $\pm$   0.4 &  35.3 $\pm$   0.7 & 0.361 $\pm$ 0.002 &  11.3 $\pm$   0.2 &  24.0 $\pm$   0.8 & 495.3 &294 \\
 &       4 &3.221 & 146.0 $\pm$   3.9 & 232.0 $\pm$   0.0 &  28.9 $\pm$   1.2 &  10.2 $\pm$   0.3 & 249.2 $\pm$   0.2 &  28.3 $\pm$   0.4 & 0.360 $\pm$ 0.001 &   9.1 $\pm$   0.1 &  19.3 $\pm$   0.4 & 619.8 &365 \\
{060526s\footnote{Second flare detected by the XRT.}} &     tot &3.221 & 148.9 $\pm$   4.0 & 282.3 $\pm$   1.0 &  18.8 $\pm$   2.4 &  36.7 $\pm$   0.8 & 308.6 $\pm$   1.5 &  72.1 $\pm$   0.8 & 0.509 $\pm$ 0.002 &  17.7 $\pm$   0.7 &  54.4 $\pm$   0.6 &1024.5 &948 \\
 &       1 &3.221 &  44.1 $\pm$   3.1 & 284.6 $\pm$   3.0 &  33.3 $\pm$   8.0 &  41.5 $\pm$   1.8 & 321.8 $\pm$   3.8 &  88.8 $\pm$   1.9 & 0.467 $\pm$ 0.005 &  23.7 $\pm$   1.6 &  65.2 $\pm$   1.3 & 823.0 &642 \\
 &       2 &3.221 &  84.6 $\pm$   2.6 & 285.6 $\pm$   1.3 &  16.1 $\pm$   3.1 &  30.8 $\pm$   0.9 & 307.9 $\pm$   1.9 &  60.7 $\pm$   0.9 & 0.507 $\pm$ 0.004 &  15.0 $\pm$   0.9 &  45.7 $\pm$   0.7 &1306.2 &850 \\
 &       3 &3.221 &  27.3 $\pm$   1.4 & 288.9 $\pm$   1.0 &   4.3 $\pm$   1.7 &  28.2 $\pm$   1.5 & 299.9 $\pm$   2.0 &  45.1 $\pm$   1.6 & 0.625 $\pm$ 0.015 &   8.5 $\pm$   1.2 &  36.7 $\pm$   1.3 & 495.3 &294 \\
 &       4 &3.221 &  26.3 $\pm$   1.5 & 288.4 $\pm$   0.9 &   2.2 $\pm$   1.0 &  27.2 $\pm$   0.0 & 296.1 $\pm$   1.8 &  39.8 $\pm$   0.4 & 0.684 $\pm$ 0.001 &   6.3 $\pm$   1.2 &  33.5 $\pm$   1.2 & 619.8 &365 \\
060904B &tot &0.703 & 392.0 $\pm$   5.2 & 124.4 $\pm$   0.9 &  72.4 $\pm$   5.4 &  29.5 $\pm$   0.6 & 170.6 $\pm$   1.3 &  79.5 $\pm$   1.5 & 0.371 $\pm$ 0.004 &  25.0 $\pm$   0.4 &  54.5 $\pm$   0.4 & 872.3 &771 \\
 &       1 &0.703 &  47.6 $\pm$   1.1 & 112.0 $\pm$   4.6 & 176.0 $\pm$  34.0 &  36.5 $\pm$   2.6 & 192.1 $\pm$   5.1 & 114.2 $\pm$   9.4 & 0.320 $\pm$ 0.029 &  38.8 $\pm$   1.0 &  75.3 $\pm$   2.3 & 445.3 &405 \\
 &       2 &0.703 & 169.9 $\pm$   2.6 & 114.0 $\pm$   1.9 & 160.0 $\pm$  17.0 &  25.0 $\pm$   0.8 & 177.2 $\pm$   2.5 &  83.3 $\pm$   5.7 & 0.300 $\pm$ 0.014 &  29.2 $\pm$   0.5 &  54.1 $\pm$   0.6 & 907.1 &875 \\
 &       3 &0.703 &  75.7 $\pm$   1.7 & 123.1 $\pm$   2.3 &  92.0 $\pm$  16.0 &  21.4 $\pm$   1.0 & 167.5 $\pm$   2.9 &  65.2 $\pm$   5.0 & 0.328 $\pm$ 0.013 &  21.9 $\pm$   0.7 &  43.3 $\pm$   0.7 & 408.0 &372 \\
 &       4 &0.703 & 114.1 $\pm$   2.4 & 123.1 $\pm$   1.5 &  92.0 $\pm$  12.0 &  17.0 $\pm$   0.6 & 162.7 $\pm$   1.9 &  54.6 $\pm$   3.9 & 0.312 $\pm$ 0.009 &  18.8 $\pm$   0.4 &  35.8 $\pm$   0.5 & 527.4 &437 \\
060929 &tot & -1.0 &  68.6 $\pm$   0.9 & 410.7 $\pm$   5.9 & 290.0 $\pm$  49.0 &  47.5 $\pm$   1.6 & 528.1 $\pm$   8.1 & 156.7 $\pm$  18.1 & 0.303 $\pm$ 0.011 &  54.6 $\pm$   1.9 & 102.1 $\pm$   1.2 & 652.5 &517 \\
 &       1 & -1.0 &  14.5 $\pm$   0.4 & 410.0 $\pm$   0.0 & 366.0 $\pm$  21.0 &  53.4 $\pm$   2.4 & 549.8 $\pm$   1.8 & 180.9 $\pm$   4.7 & 0.295 $\pm$ 0.005 &  63.7 $\pm$   1.0 & 117.1 $\pm$   3.3 & 278.6 &226 \\
 &       2 & -1.0 &  29.3 $\pm$   0.6 & 410.0 $\pm$   0.0 & 306.0 $\pm$  14.0 &  46.7 $\pm$   1.3 & 529.5 $\pm$   1.4 & 156.6 $\pm$   3.4 & 0.298 $\pm$ 0.003 &  54.9 $\pm$   0.5 & 101.6 $\pm$   1.7 & 470.6 &382 \\
 &       3 & -1.0 &  11.9 $\pm$   0.4 & 432.0 $\pm$  13.0 & 157.0 $\pm$  81.0 &  46.2 $\pm$   3.7 & 517.2 $\pm$  19.0 & 133.7 $\pm$  28.9 & 0.346 $\pm$ 0.016 &  43.7 $\pm$   5.6 &  89.9 $\pm$   3.3 & 143.5 &135 \\
 &       4 & -1.0 &  17.4 $\pm$   0.5 & 436.0 $\pm$  10.0 & 130.0 $\pm$  58.0 &  46.0 $\pm$   2.8 & 513.3 $\pm$  15.3 & 127.8 $\pm$  20.8 & 0.360 $\pm$ 0.012 &  40.9 $\pm$   4.8 &  86.9 $\pm$   3.1 & 263.2 &187 \\
070520B &tot & -1.0 &  78.7 $\pm$   1.4 & 146.6 $\pm$   1.8 &  55.0 $\pm$   7.6 &  37.9 $\pm$   1.2 & 192.3 $\pm$   2.5 &  91.4 $\pm$   1.8 & 0.415 $\pm$ 0.006 &  26.8 $\pm$   0.9 &  64.7 $\pm$   0.9 & 480.2 &431 \\
 &       1 & -1.0 &  21.4 $\pm$   0.6 &  92.0 $\pm$  11.0 & 420.0 $\pm$ 130.0 &  31.9 $\pm$   2.9 & 207.8 $\pm$  13.1 & 125.6 $\pm$  50.2 & 0.254 $\pm$ 0.137 &  46.9 $\pm$   2.0 &  78.8 $\pm$   2.0 & 265.6 &262 \\
 &       2 & -1.0 &  39.1 $\pm$   1.0 & 134.1 $\pm$   3.2 & 104.0 $\pm$  19.0 &  32.0 $\pm$   1.6 & 191.8 $\pm$   4.1 &  91.7 $\pm$   5.6 & 0.349 $\pm$ 0.014 &  29.8 $\pm$   1.1 &  61.8 $\pm$   1.2 & 460.4 &365 \\
 &       3 & -1.0 &  12.0 $\pm$   0.6 & 152.7 $\pm$   4.0 &  31.0 $\pm$  16.0 &  30.7 $\pm$   2.9 & 183.6 $\pm$   6.8 &  68.8 $\pm$   3.7 & 0.446 $\pm$ 0.014 &  19.0 $\pm$   2.6 &  49.7 $\pm$   2.0 & 137.2 &93 \\
 &       4 & -1.0 &  13.4 $\pm$   0.8 & 164.3 $\pm$   2.1 &   8.1 $\pm$   4.5 &  33.9 $\pm$   2.7 & 180.9 $\pm$   4.1 &  58.3 $\pm$   2.7 & 0.582 $\pm$ 0.019 &  12.2 $\pm$   2.3 &  46.1 $\pm$   2.1 & 131.5 &102 \\
070704 &tot & -1.0 &  85.4 $\pm$   1.3 & 238.9 $\pm$   3.5 & 201.0 $\pm$  30.0 &  34.6 $\pm$   1.7 & 322.3 $\pm$   4.3 & 112.9 $\pm$   9.5 & 0.307 $\pm$ 0.012 &  39.1 $\pm$   0.8 &  73.7 $\pm$   1.3 & 537.2 &498 \\
 &       1 & -1.0 &   1.8 $\pm$   0.2 & 220.0 $\pm$   0.0 & 300.0 $\pm$   0.0 &  41.8 $\pm$   3.5 & 332.0 $\pm$   4.6 & 143.1 $\pm$   9.1 & 0.292 $\pm$ 0.010 &  50.6 $\pm$   2.8 &  92.4 $\pm$   6.3 &   8.9 &20 \\
 &       2 & -1.0 &  25.4 $\pm$   0.7 & 256.9 $\pm$   3.7 &  84.0 $\pm$  18.0 &  54.0 $\pm$   2.5 & 324.2 $\pm$   5.9 & 132.1 $\pm$   4.6 & 0.409 $\pm$ 0.007 &  39.1 $\pm$   2.0 &  93.1 $\pm$   1.8 & 370.7 &316 \\
 &       3 & -1.0 &  20.7 $\pm$   0.6 & 250.7 $\pm$   4.3 & 104.0 $\pm$  24.0 &  44.7 $\pm$   2.5 & 318.9 $\pm$   6.2 & 119.1 $\pm$   6.6 & 0.375 $\pm$ 0.009 &  37.2 $\pm$   1.9 &  81.9 $\pm$   1.8 & 308.3 &240 \\
 &       4 & -1.0 &  38.6 $\pm$   0.9 & 249.7 $\pm$   3.5 & 104.0 $\pm$  22.0 &  38.2 $\pm$   2.3 & 312.7 $\pm$   5.0 & 105.3 $\pm$   5.7 & 0.363 $\pm$ 0.009 &  33.6 $\pm$   1.3 &  71.8 $\pm$   1.7 & 410.4 &362 \\
090621A &tot & -1.0 & 325.8 $\pm$   4.9 & 203.1 $\pm$   2.3 & 116.0 $\pm$  16.0 &  21.1 $\pm$   0.6 & 252.6 $\pm$   2.7 &  68.0 $\pm$   5.7 & 0.310 $\pm$ 0.007 &  23.5 $\pm$   0.7 &  44.6 $\pm$   0.4 &1006.6 &636 \\
 &       1 & -1.0 &    -1.0 &   -1.0 &    -1.0 &    -1.0 &  -1.0 &    -1.0 &  -1.0 &    -1.0 &    -1.0 &    -1.0 & -1.0 \\
 &       2 & -1.0 &  53.2 $\pm$   1.4 & 197.5 $\pm$   4.5 & 156.0 $\pm$  30.0 &  24.1 $\pm$   1.4 & 258.8 $\pm$   4.2 &  80.6 $\pm$   9.9 & 0.299 $\pm$ 0.014 &  28.2 $\pm$   0.8 &  52.3 $\pm$   1.2 & 380.8 &346 \\
 &       3 & -1.0 &  69.7 $\pm$   1.7 & 199.1 $\pm$   4.8 & 150.0 $\pm$  37.0 &  20.5 $\pm$   1.2 & 254.6 $\pm$   5.4 &  70.5 $\pm$  13.8 & 0.291 $\pm$ 0.017 &  25.0 $\pm$   1.1 &  45.5 $\pm$   0.8 & 452.0 &375 \\
 &       4 & -1.0 & 194.1 $\pm$   3.6 & 199.3 $\pm$   3.2 & 144.0 $\pm$  26.0 &  18.5 $\pm$   0.7 & 251.0 $\pm$   3.8 &  64.6 $\pm$   9.8 & 0.290 $\pm$ 0.012 &  23.0 $\pm$   0.8 &  41.6 $\pm$   0.5 &1000.3 &726 \\
 \hline
  \label{Tab:bestfit}
 \end{tabular}}
  \end{minipage}
\end{table*}

\begin{table*}
  \centering
  \begin{minipage}{180mm}
    \caption{Best fit parameters derived from the spectral modelling of the XRT and BAT data of GRB\,060904B. A photo-electrically absorbed models
    (\textsc{tbabs*ztbabs} within \textsc{xspec}). Two different spectral models have been used: a simple power-law (SPL) and a band function with
    the peak of the $\nu  F_{\nu}$ spectrum as free parameter. From left to right: name of the interval of the extraction of the spectrum: XRT+BAT
    stands for a joint BAT-XRT data fitting; start and stop time of extraction of each spectrum; model used; intrinsic neutral Hydrogen column density;
    best fit low and high energy photon indices for a Band function or best fit photon index $\Gamma$ for a SPL model; spectral peak energy; 
    isotropic luminosity; statistical information about the fit. The $^{*}$ symbol indicates
    an apparent trend in the residuals of the fit. A joint fit of all the spectra gives $N_{\rm{H,z}}=0.5\pm0.1$, $\chi^2/$dof=$579.76/567$, 
    P-value=35\%.The Galactic absorption has been frozen to $1.13\times10^{21}\,\rm{cm^{-2}}$ (\citealt{Kalberla05}). Errors are provided at 90\% c.l.}
  \resizebox{\textwidth}{!}{
    \begin{tabular}{ccccccccccccc}
      \hline
    &  Interval  & $t_{i}$ & $t_{f}$  & Model  & $N_{\rm{H,z}} $         & $\alpha_{\rm{B}}$ & $\beta_{\rm{B}}(\Gamma)$  & $E_{\rm{peak}}$  &    $L_{\rm{iso}}$     &$\chi^2$/dof  & P-value  \\
   &            &  (s)    &  (s)    &        &($10^{22}\,\rm{cm^{-2}}$)&                  &                          &   (keV)         & ($10^{50}$ erg $\rm{s}^{-1}$)&       &\\
      \hline
    1 & XRT+BAT  &129      &150      &SPL     &$2.71\pm 0.11$          &       --         &$1.85\pm0.50$             & --   &--      &  $58.60/45$   &  $8\times 10^{-2}$& $^{*}$\\  
      &          &         &         &Band    &$0.52$           & $0.49\pm0.14$    &$2.60$             &$7.2^{+1.1}_{-0.9}$ &$0.13\pm0.01$& $56.99/45$   &  0.11     \\
      \hline
    2 & XRT+BAT  &150      &165      &SPL     &$2.52\pm0.30$           &       --         &$2.09\pm0.05$             & --     &--    &  $166.81/101$ &  $4\times 10^{-5}$& $^{*}$\\
      &          &         &         &Band    &$0.52$           & $0.50$           &$2.50\pm0.15$             &$5.9_{-0.5}^{+0.6}$&$0.82\pm0.08$& $111.48/101$ &      0.22    \\
      \hline 
   3 & XRT+BAT  &165      &176      &SPL     &$2.56\pm0.30$           &       --         &$2.24\pm0.05$             & --     & --   &  $161.81/91$  &  $7\times 10^{-6}$& $^{*}$\\
      &          &         &         &Band    &$0.52$           & $0.77$          &$2.60\pm0.25$             &$4.7_{-0.6}^{+0.8}$& $0.82\pm0.08$&$76.20/90$&     0.85\\
      \hline 
   4 & XRT+BAT  &176      &190      &SPL     &$1.38\pm0.20$           &       --         &$2.27\pm0.07$             & --     & --   &  $105.28/85$  &  $7\times 10^{-2}$& $^{*}$\\
      &          &         &         &Band    &$0.52$           & $1.00$          &$2.59\pm0.21$             &$ 3.2_{-0.3}^{+0.3}$  &$0.54\pm0.05$&$76.13/85$&     0.74    \\
      \hline
   5 & XRT+BAT  &190      &210      &SPL     &$1.06\pm0.15$           &       --         &$2.46\pm0.10$             & --      & --  &  $64.33/88$   &  $0.97$\\
      &          &         &         &Band    &$0.52$           & $1.00$          &$2.59\pm0.16$             &$1.7_{-0.2}^{+0.2}$&$0.35\pm0.03$& $63.50/88$&     0.98\\
      \hline 
  6 & XRT+BAT  &210      &224      &SPL     &$0.82\pm0.14$           &       --         &$2.69\pm0.13$             & --      &  -- &  $89.96/78$   &  $0.17$\\ 
      &          &         &         &Band    &$0.52$           & $1.00$         &$2.72\pm0.14$&             $<1.01$     & $0.17\pm0.02$ &$ 85.34/78$   &  0.27\\
      \hline 
  7 & XRT       &224      &249      &SPL     &$0.88\pm0.12$           &       --         &$3.05\pm0.13$             & --     &--   &  $79.94/78$   &  $0.42$\\
      &          &         &         &Band    &$0.52$           & $1.00$         &$3.05\pm0.22$             &$<0.4$& $0.087\pm0.003$ &$91.89/77$  &   0.12 \\
      \hline
   8 & XRT      &249      &320      &SPL     &$0.71\pm0.09$           &       --         &$3.60\pm0.18$             & --       &--  &  $92.80/70$   &  $0.04$\\
     &          &         &         &SPL     &$0.52$                  &       --         &$3.28\pm0.08$             & --      &$0.030\pm0.003$&  $105.94/71$   & 0.01\footnote{The presence 
of a spectral feature in excess of the SPL component in this interval of time is discussed in \cite{Margutti08} and \cite{Moretti08}.}\\
  \hline
\label{Tab:spec060904B}
  \end{tabular}}
  \end{minipage}
\end{table*}

\begin{table*}
  \centering
  \begin{minipage}{180mm}
    \caption{Same as Table \ref{Tab:spec060904B} for GRB\,060418. A joint fit of all the spectra gives $N_{\rm{H,z}}=0.6\pm0.1$, $\chi^2/$dof=$318.43/327$, P-value=62\%.
    The Galactic absorption has been frozen to $8.81\times10^{20}\,\rm{cm^{-2}}$ (\citealt{Kalberla05}). Errors are provided at 90\% c.l.}
  \resizebox{\textwidth}{!}{
    \begin{tabular}{ccccccccccccc}
      \hline
    &  Interval  & $t_{i}$ & $t_{f}$  & Model  & $N_{\rm{H,z}} $         & $\alpha_{\rm{B}}$ & $\beta_{\rm{B}}(\Gamma)$  & $E_{\rm{peak}}$  &  $L_{\rm{iso}}$ & $\chi^2$/dof  & P-value  \\
   &            &  (s)    &  (s)    &        &($10^{22}\,\rm{cm^{-2}}$)&                  &                          &   (keV)    &        ($10^{50}$ erg $\rm{s}^{-1}$)        &\\
      \hline
  1 & XRT+BAT   &122   &131  &SPL     &$2.2\pm0.50$& --        &$2.0\pm0.1$  & --                     &--&  $126.03/64$&   $6\times10^{-6}$& $^{*}$\\
    &           &      &     &Band    &$0.59$      & 0.90      &$2.37\pm0.22$&$5.1_{-0.7}^{+0.9}$ &$4.2\pm0.4$&  $74.26/64$&      0.18\\
  \hline
  2 & XRT+BAT   &131   &140  &SPL     &$2.6\pm0.4$ & --        &$2.1\pm0.1$  & --              &--&             $143.20/103$&  $5\times10^{-3}$\\ 
    &           &      &     &Band    &$0.59$      & 0.90      &$2.33\pm0.13$&$4.57_{-0.6}^{+0.5}$ &$6.6\pm0.6$& $86.16/103$&      0.88\\
  \hline
  3 & XRT+BAT   &140   &148  &SPL     &$1.91\pm0.42$& --       &$2.29\pm0.12$&  --              &--&              $62.85/58$&      0.31\\
    &           &      &     &Band    &$0.59$      & 1.00      &$2.50\pm0.40$&$2.2_{-0.4}^{+0.3}$ &$2.6\pm0.2$& $62.30/58$&      0.33\\
  \hline
  4 & XRT+BAT   &148   &170  &SPL     &$1.18\pm0.17$& --       &$2.44\pm0.09$&  --              &--&              $96.44/107$&      0.76\\
    &           &      &     &Band    &$0.59$      & 1.00      &$2.44\pm0.11$&$1.18_{-0.1}^{+0.14}$&$1.0\pm0.1$&  $96.32/107$&      0.76\\
  \hline
\label{Tab:spec060418}
  \end{tabular}}
  \end{minipage}
\end{table*}

\label{lastpage}
\end{document}